# The Influencer Next Door: How Misinformation Creators Use GenAI

*Amelia Hassoun[1], Ariel Abonizio, Katy Osborn, Cameron Wu, Beth Goldberg*

# Introduction

Academic literature on generative AI (GenAI) has largely focused on the problem of detecting and discerning AI-generated content from human-generated content (Caldwell et al., 2015). This literature often distinguishes expert creators, like state-sponsored disinformation spreaders, from everyday consumers, framed as non-experts who passively see GenAI content and must discern whether it is 'real' or 'fake.'

Based on a longitudinal ethnographic study of misinformation creators and consumers, we find this distinction increasingly untenable. For participants, GenAI tools largely served not as an oracle for (mis)information discovery or truth, but as an aid for bricolage work (Levi-Strauss, 1966)—remixing and redeploying existing content to meet their needs. Our findings evidence a need to shift analysis from the public as consumers of AI content to bricoleurs who creatively use GenAI.

Our research yielded four key findings:

> **1. GenAI for Creation, not Truth Discernment:** Few participants used GenAI as a truth-seeking tool. Rather, they used it for content creation**.**
>
> **2. Motivated by Money and Productivity:** A growing belief in the 'influencer millionaire' narrative drove participants to become content creators. Participants used GenAI primarily as a productivity tool to generate volumes of (often misinformative) content.
>
> **3. Democratization of (Mis)information Creation:** GenAI lowered the barrier to entry for online content creation, enabling participants to produce and share content without formal technology or marketing training.
>
> **4. Use of Influencer Marketing Tactics:** Creators learned and deployed influencer marketing tactics to expand engagement and monetize content.

---

[1] Corresponding Author. <u>ah2229@cam.ac.uk</u>. Darwin College, University of Cambridge. Silver St, CB3 9EU.

Our findings suggest that financial motivations play a major role in turning misinformation consumers into creators. While misinformation creation has long been financialized (Ayeb and Bonini, 2024), we argue three factors are likely accelerating financial motives behind misinformation production.

First, people increasingly see online content creation as a reputable, desirable, and accessible path to wealth. Many participants began because they believed anyone could become rich by creating engaging content online. He et al. (2023) found 57% of 13-26 year-olds surveyed wanted to become an influencer, 53% saw it as a reputable career choice, and 50% would quit their jobs if they could earn enough. While influencer studies often survey younger generations, we found this attitude present across generations.

Second, aspiring creators made misinformation content because they discovered that social media algorithms rewarded the engagement it generated, leading to monetization opportunities. Though no participant expressed intent to create misinformation, this incentive meant creators cared more about virality than veracity. Misinformation creators did not think of themselves as malicious actors: they saw themselves as individuals leading careers as influencers and entertainers.

Third, GenAI tools have democratized both content creation and monetization. For creator participants, experiments with GenAI confirmed anyone could be a content creator, even without writing, video editing, or coding skills. GenAI tools made creating misinformation at the speed and volume needed to monetize increasingly accessible. This motivated participants to transition from consumers and hobbyist creators to entrepreneurial micro-influencers learning how to use GenAI for lucrative self-marketing.

We define 'misinformation' as 'ideas that have yet to be the subject of a strong consensus of experts' (Uscinski 2023:11) and employ Krause et al. (2022)'s framework for evaluating misinformation threat vis-à-vis science-related truth claims. Despite its limitations, we use the term because participants articulated how they experienced (and evaded) moderation based upon platform misinformation policies and because research evidences harms from content advocating practices that contravene strong expert consensus (Perlis et al., 2022).

Next, we situate our work in existing literature and detail our methodology. We then discuss our findings and their implications for research on GenAI's effects.

## Focus on Information- and Truth-Seeking Behavior

The most widely-cited literature analyzing AI-driven misinformation focuses on AI models' unreliability, demonstrating how they produce erroneous information due to training data, biases, and other design factors (Noble, 2018; Crawford, 2021). These studies portray users as information-seekers often misled by AI and/or negatively affected by its judgments.

Studies on user interaction with GenAI largely focus on whether people can discern 'fake' (i.e. AI-generated) from 'real' (i.e. human-generated) content when seeking or encountering information online. For example, Jin et al. (2023) analyze users' credibility assessments of deepfake videos, experimenting with content, context (e.g. video popularity, source cues) and length. These works assess the negative effects of AI-generated misinformation, studying their impact on political attitudes (Dobber et al., 2021), democratic elections (Diakopoulos and Johnson, 2020), and trust (Abadie et al., 2024; Liu and Wang, 2024, Scholz et al., 2024). Suggested interventions involve increasing user data literacy (Fotopoulou, 2020), social learning (McCosker, 2022), human AI detection abilities (Hargittai et al, 2020), and automated labeling (Zhou et al., 2023).

While useful, their paradigm of information-seeking and assessment covers only a subset of GenAI uses—and, we argue, potentially excludes more common and harmful uses. We instead focus on how people use GenAI tools to *create* (mis)information, analyzing their motivations, behaviors, and the effects of their creations.

## Democratization of Creation

Organized weaponization of AI for misinformation dissemination has become a pressing academic and policy concern. Researchers contend that AI-powered misinformation campaigns, deepfakes, and autonomous weapons pose significant harms, given their potential to manipulate public opinion, sow discord, and disrupt democratic institutions (Brundage et al., 2018).

Literature on AI-generated misinformation often focuses on organized and/or state-sponsored campaigns. Georgetown's Center for Security and Emerging Technology (2021) analyzes how GenAI boosts state-sponsored disinformation campaigns by facilitating increased production and quality.

Misinformation scholars often define misinformation as verifiably false information (Dan et al., 2021). Consequently, analysis of GenAI's role in misinformation production and distribution has largely focused on high-quality synthetic images and videos of fabricated events. This scholarship argues that the increased creation of targeted, high-quality deepfakes is GenAI's most significant contribution to increased misinformation belief among individuals (Bontridder and Poullet, 2021; Westerlund, 2019).

Instead, we show how GenAI is (mis)used not only by organized experts like politically-motivated state-sponsored actors, but by individual creators seeking monetizable engagement. Emerging technologies increasingly democratize misinformation production (Diaz Ruiz, 2023). GenAI tools significantly accelerate it. We argue that existing literature currently under-emphasizes GenAI's use as a personal productivity tool, increasing the quantity and speed of (mis)information creation and sharing. Aspiring micro-influencers intentionally focus on quantity over quality to drive engagement (Hassoun et al., 2024), and GenAI tools accelerate their capacity to generate that quantity.

We argue that the increased *quantity* of misinformation produced by non-experts with GenAI could pose more harm than high-quality fakes (Paris and Donovan, 2019). GenAI's ability to facilitate replication of misleading 'bricolage' content (Hassoun et al., 2024) at scale creates different moderation challenges than blatant falsehoods and quality deepfakes more easily detected and removed by existing platform policies (Baker et al., 2020).

**Financial Motives for Disinformation Production**

Between the US General Election in 2016 and GenAI's advent, misinformation creators adopted mainstream news aesthetics to gain trust (Keener, 2018). Our research evidences another trust-manufacturing strategy, enabled by GenAI tools: misinformation creators increasingly define themselves as influencers, borrowing those communities' tactics and aesthetics. Further, trust in influencers is growing: the percentage of 13-42 year-olds who trust social media influencers grew from 51% in 2019 to 61% in 2023 (He et al., 2023).

After January 6th, studies investigated individual participation in misinformation ecosystems (Prochaska et al., 2023), finding many individual creators spread false or misleading claims in search of profit (Freelon and Wells, 2020). We found similar financial motivations fueling GenAI use for misinformation creation. Participants saw becoming an influencer as a reputable, desirable, and accessible path to wealth. This 'influencer millionaire' narrative was popularized by people like 18-year-old Charli D'Amelio, who gained 150 million followers by creating lifestyle content (Fetter et al., 2023). Influencer idolization is well-documented in marketing literature, but requires interdisciplinary research as it inspired GenAI misinformation creators' actions.

Despite microinfluencers' small following, they substantially impact misinformation spread by manufacturing familiarity, relatability, and authenticity which motivates trusted sharing (Anspach, 2017; Harff et al., 2022; Stehr et al., 2015). We find micro-influencers professionalizing by leveraging and teaching marketing tactics. GenAI tools democratize the ability to learn and deploy these tactics, lowering barriers to entry and raising ceilings of financial opportunity for aspiring content creators.

In short, we suggest that the dominant harm narrative in GenAI and misinformation research—that 'ordinary' people are duped by AI-generated content created by 'experts' or state actors—is incomplete. We argue GenAI makes misinformation creation less the sole purview of organized political actors and more a tool for everyday users to increase engagement and, by extension, profit. For creators, belief in misinformation content was often unimportant or secondary to generating capital from content, requiring corresponding analytical and platform approaches to counteracting AI-driven misinformation spread.

# Methodology

We conducted a 15-month ethnographic study in Brazil and the United States between 2022-2023 to longitudinally analyze why and how people consume, amplify and create political and medical misinformation.[2] During this research, ChatGPT was released, leading us to analyze GenAI's early impacts on misinformation creation, amplification, and consumption.

We employed ethnographic methods to contextually understand participants' content creation motivations and practices. We observed in situ how creators used GenAI to increase, automate, and professionalize creation; and the tactics and trust heuristics they learned and deployed to elicit engagement.

---

[2] See Hassoun et al. 2024 for detailed methodology.

## Participants & Sites

In 2022 we conducted ethnographic research with 31 participants aged 18-67 who regularly created, amplified, and/or consumed misinformation (Table 1). To recruit, we identified relevant social media and chat app groups and disclosed ourselves as researchers.

**Table 1**

*Phase 1 Participant Information*

|  | Total N | Men (self-ID) | Women (self-ID) | Urban | Rural | Misinformation Type | | |
|---|---|---|---|---|---|---|---|---|
|  |  |  |  |  |  | Medical (only) | Political (only) | Both |
| Site |  |  |  |  |  |  |  |  |
| Brazil | 16 | 9 | 7 | 12 | 4 | 2 | 4 | 10 |
| US | 15 | 6 | 9 | 10 | 5 | 5 | 4 | 6 |
| **TOTAL** | 31 | 15 | 16 | 22 | 9 | 7 | 8 | 16 |

In 2023, we replicated the method with 25 participants (Table 2). 15 participants were Phase 1 re-contacts. 4 used GenAI tools. We recruited 10 new participants who regularly used GenAI. Most utilized more than one type of GenAI tool (e.g. voice *and* text synthesis). To recruit, we identified misinformation in Phase 1 participants' online communities that was publicly fact-checked or we suspected was created using GenAI, and contacted creators.

**Table 2**

*Phase 2 Participant Information*

|  | Total N | Men (self-ID) | Women (self-ID) | Urban | Rural | Misinformation Type | | |
|---|---|---|---|---|---|---|---|---|
|  |  |  |  |  |  | Medical (only) | Political (only) | Both |
| Site |  |  |  |  |  |  |  |  |
| Brazil | 12 | 8 | 4 | 9 | 3 | 2 | 2 | 8 |
| US | 13 | 7 | 6 | 8 | 5 | 5 | 3 | 5 |
| **TOTAL** | 25 | 15 | 10 | 17 | 8 | 7 | 5 | 12 |

**Table 3**

*Participant use of GenAI by category*

|  |  | GenAI Use by category | | | |
|---|---|---|---|---|---|
|  | Total N | Text | Image | Video | Voice |
| Site |  |  |  |  |  |
| Brazil | 4 | 4 | 1 | 1 | 2 |
| US | 7 | 4 | 5 | 2 | 0 |
| **TOTAL** | 11 | 8 | 6 | 3 | 2 |

Our study passed human subjects ethics review and participants gave informed consent, receiving $100/hour (US) or 250 reais/hour (Brazil). Researchers natively spoke English and Portuguese. All personally identifying information is omitted. Learning from participants, we used GenAI tools to re-generate similar visual and written content artifacts to prevent re-identification.

## Research Methods

In Phase 1, we conducted three-part, 6-8 hour ethnographic interviews with each participant: 1 online semi-structured interview, 1 in-person (in participant homes and social spaces) semi-structured interview and screen-sharing observation, and 1 semi-structured interview with a secondary participant (important to and introduced by the primary participant). We also conducted participant-observation at 3 misinformation-spreading events and qualitatively analyzed 30+ English- and Portuguese-speaking misinformation-spreading communities across social media platforms throughout the study. In Phase 2, we used the following methods:

*'Re-contacts' (n=15): Misinformation Consumers, Amplifiers & Creators*

We conducted 2-4 hours of semi-structured interview and participant-observation over 1-2 sessions to analyze belief, behavior, and misinformation ecosystem evolution and GenAI awareness, perceptions, and use.

*New Participants (n=10): Misinformation Creators Using GenAI*

We conducted online 1-2 hour semi-structured interviews followed by 4-8 hours of in-person[3] interviews and participant-observation over 1-3 sessions.

*GenAI Prompting Exercise (n=25)*

We prompted all participants with standardized AI-generated misinformation images and videos (without revealing they were AI-generated), accompanied by a structured interview examining participants' ability to recognize AI-generated misinformation, perceived everyday exposure, and GenAI's effect on their information navigation, content-sharing, and trust heuristics. We then revealed that content was AI-generated to observe and discuss their reactions.

## Analysis Methods

Researchers recorded images, video, and notes during participant-observation. We conducted grounded theory-guided data analysis (Charmaz, 2006), collaboratively performing open, clustering, and thematic coding (Saldaña, 2021).

## Limitations

Ethnographic methods facilitated in-depth observational, qualitative analysis of phenomena usually studied with lab-based experiments or computational models (Seo and Faris, 2021). Sampling bias may affect GenAI misinformation creator findings, as participants either disclosed their GenAI use, created publicly fact-checked content, or had researcher-discernible AI use. A large-scale survey would help validate ethnographic findings with a representative sample.

Self-reported data suffers from self-censoring, recall, and social desirability biases. To mitigate, we cross-referenced data from semi-structured interviews with digital artifacts (e.g., search and message histories) and screen-sharing observations. This allowed us to analyze discrepancies between participants' self-reported and actual behaviors.

---

[3] We conducted online-only interviews and participant-observation with 5 respondents who presented potential risk to researcher safety.

# Findings

We first explain how and why participants used GenAI for creation rather than truth discernment, demonstrating how becoming a productive influencer 'millionaire next door' motivated creators. Second, we explain how GenAI tools democratize misinformation creation, showing how creators learned from them. Third, we detail how creators used marketing tactics to expand engagement and monetize content.

## Motivations for GenAI Use

*Creation (Not Truth Discernment)*

Participants rarely encountered new misinformation on GenAI tools or used them to search for misinformation, because queries seldom generated satisfying answers. For example, when Enrique (43, BR) asked ChatGPT about 'Q Drops', he received a generic greeting message unrelated to QAnon. Rogerio (22, BR) had similar null results when asking Bard[4] about 'the medical benefits of adrenochrome', getting a broad chemical compound description stating no known medical benefits. Both Enrique and Rogerio wanted to validate pre-existing beliefs and expected GenAI chatbots to behave like search engines cataloging the detailed conspiracy evidence they had found elsewhere online. Although queries prompted no misinformation warnings, both users were disappointed by results.

Instead, participants generally used GenAI as a bricolage tool to refine, repackage, and replicate existing content. They used GenAI tools to create or learn how to create content for personal, goal-oriented reasons: to make money, for school or work progression, or to increase social capital. They seldom sought knowledge for knowledge's sake, evaluating tools' answers based on their application rather than their truth-value: asking 'does this help achieve my goal?' not 'is this a true fact?'

*Money*

A growing belief in the 'influencer millionaire' narrative drove participants to become content creators—and GenAI enabled them to produce and share high volumes of engaging content without formal technology or marketing training. Aspiring microinfluencers created misinformation for financial gain.

---

[4] Now called "Gemini"

Otto (23, US), a DACA dreamer, began consuming misinformation at 12 after discovering an Infowars sticker at a local cafe. When his favorite teacher affirmed Alex Jones' conspiracy theories, Otto's interest and trust in alternative news sources increased. He discovered hyper-masculine 'self-help' YouTubers Owen Cook, Julian Blanc and Andrew Tate via internet chatrooms and YouTube-suggested videos. Inspired by Tate's financial success, Otto decided to pursue hobbyist content creation, embracing Tate's strategy of being controversial to capture attention. In 2022, he was working nights when he heard about ChatGPT and saw its application for financial gain, thinking: 'I'm not going to get rich by finding a job…The most profitable thing to do is to create YouTube videos.'

George (40, BR), a mattress salesman and political and medical misinformation creator, shared the same monetization belief:

> The plan I have is to reach the largest number of followers. I know that on TikTok there is no way to make money [just from views], I reached 7 million. But I know that you can post sales things. You can put things in your status sales, an advertisement, [so] these are my plans.

George felt apathetic about the presidential race and political belief did not motivate his creations. George's searches showed he sought money: he queried ChatGPT for the best ways to monetize YouTube channels and generate income online.

John (52, US), a campaign marketing executive, described content creation as a hunt for 'passive income'. John had nearly 150k personal social media followers, most from running a COVID misinformation group opposing mask and vaccine mandates. John used GenAI extensively, boasting about his ability to create and monetize engaging, viral content: he produced Substack accounts generating $40-80k/year.

*Productivity*

In part because of these financial motivations, misinformation creators largely treated GenAI as a productivity tool to increase speed and scale of (mis)information production.

Clodoval (42, BR), a Christian fundamentalist and book illustrator from a major city's outskirts, spent most of his time creating, sharing, and consuming anti-communist content.

The 2016 and 2022 presidential elections spiked his anti-communist creator activity. Clodoval used GenAI to prolifically create posts about election fraud and taught others to do the same through a Telegram channel for alt-right content creators. He said: 'Something that would take you 20 days to complete, you can use AI and you'll get it done in no time…It's just a tool to speed things up.' John (52, US) shared similar sentiments: 'Once ChatGPT hit, I immediately realized this could cut the time it takes to do my job in half.' Nadia (35, US) used ChatGPT to edit students' 'healer stories', increasing the paying clients she could manage.

Clodoval used ChatGPT to quickly generate 'filler' website content: 'I needed a ton of text, a ton of content. So I asked ChatGPT to do that.' He asked ChatGPT for the history of communism, quickly adapting it to his ideology by tweaking words and adding that the Nazis were communist. Clodoval also used image synthesis GenAI tools to imitate President Lula and quickly create clickbait thumbnails.

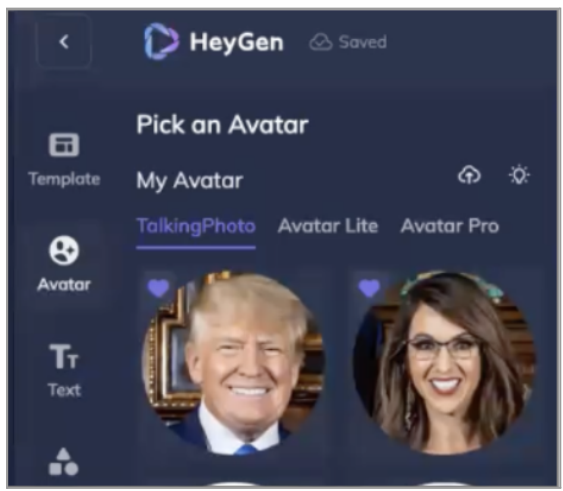

*Figure 1: Stanley (63, US) used HeyGen to create and monetize deepfake videos.*

Creators used GenAI tools to rapidly perform time-consuming simple, repetitive tasks (e.g. removing image backgrounds). Stanley (63, US) formerly needed a crew to produce content. He now produces GenAI celebrity deepfakes alone: 'I directed people on stage and in film. AI is like having a live human being…Once I've done a character, it goes into the closet and it's there as long as I have internet connection.' After encountering misinformation guardrails on mainstream GenAI tool Studio D-id preventing him from using celebrity images, he found HeyGen, an alternative without guardrails (Figure 1). Stanley also used the GenAI tool Remini to upscale his videos in dramatically less time (from hours to seconds).

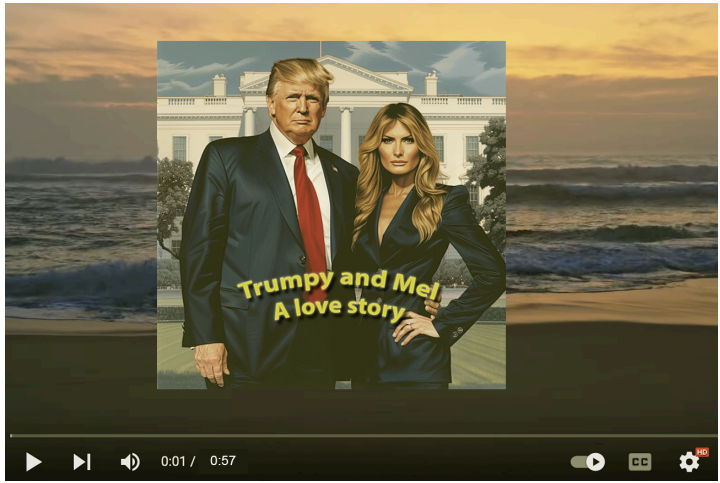

*Figure 2: Stanley's use of Remini.*

Otto (23, US) transitioned to full-time creation when he realized how productive GenAI could make him:

> I was working overnights at Amazon 10pm to 6am and then at this facility for elderly people from 3pm to 7pm…I had no free time. Then one day I was like, alright, let me see what I can do with this ChatGPT, and then I learned about MidJourney too. That's how I made [this children's book] with pictures and a cover and all that. I remember telling people the next day that it took me under 24 hours. So like, within 24 hours people were able to buy it. That's what made me get hooked.

Otto used GenAI tools to quickly generate and (re)post misinformation he read across platforms. He collected Maui fires conspiracy content from Telegram and Reddit, pasted text and links to POE.com (an LLM aggregator) to summarize the content, used PromptPerfect to create a prompt for a Maui fires conspiracy tweet-writing bot, pasted that prompt in Claude2 (via POE.com) to create the bot, and then posted the resulting AI-generated Tweets (Figure 3).

POE.com initially recognized the prompt as potentially misinformation-generating and responded 'I don't have enough information to make claims…' Yet when Otto replied 'give me 8 Tweets dude' it performed the task, illustrating how easily and routinely he bypassed misinformation guardrails.

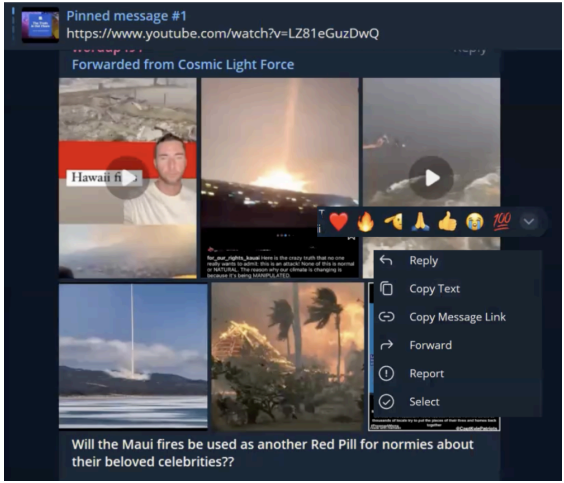
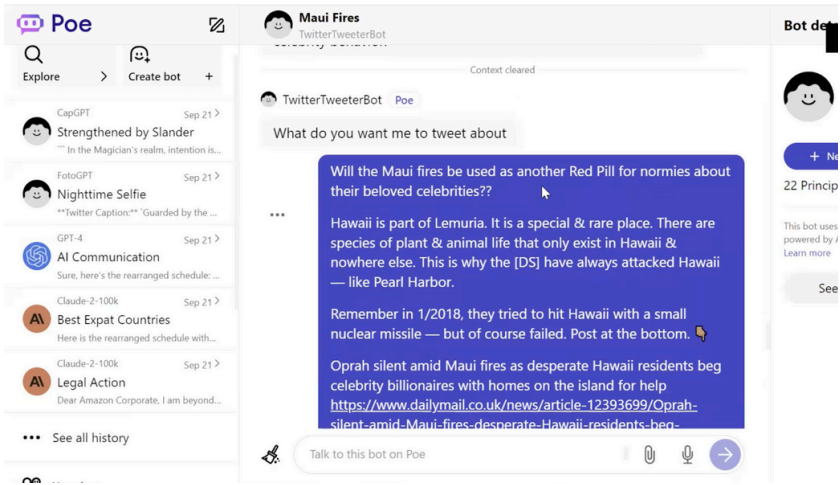
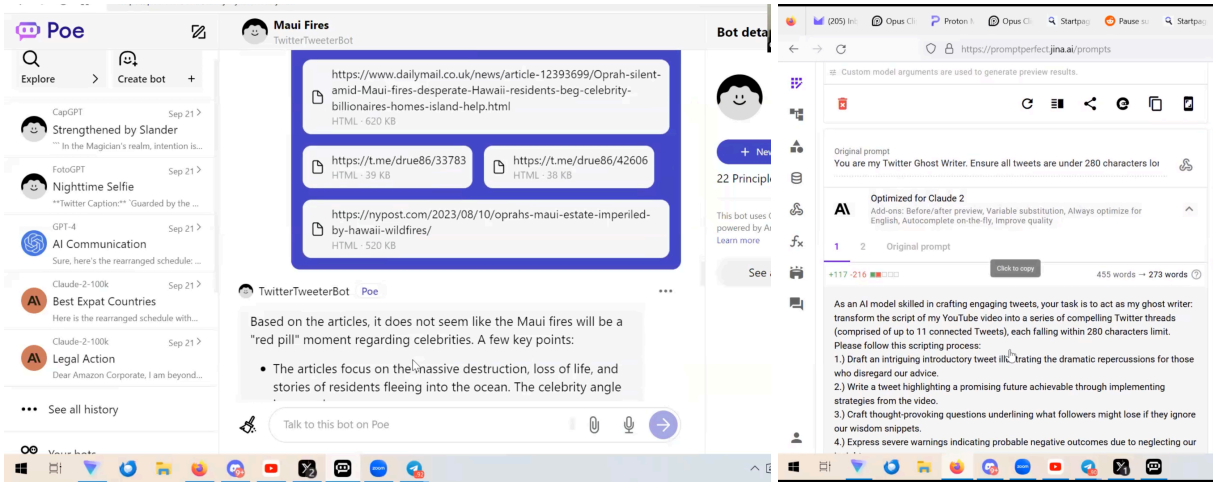

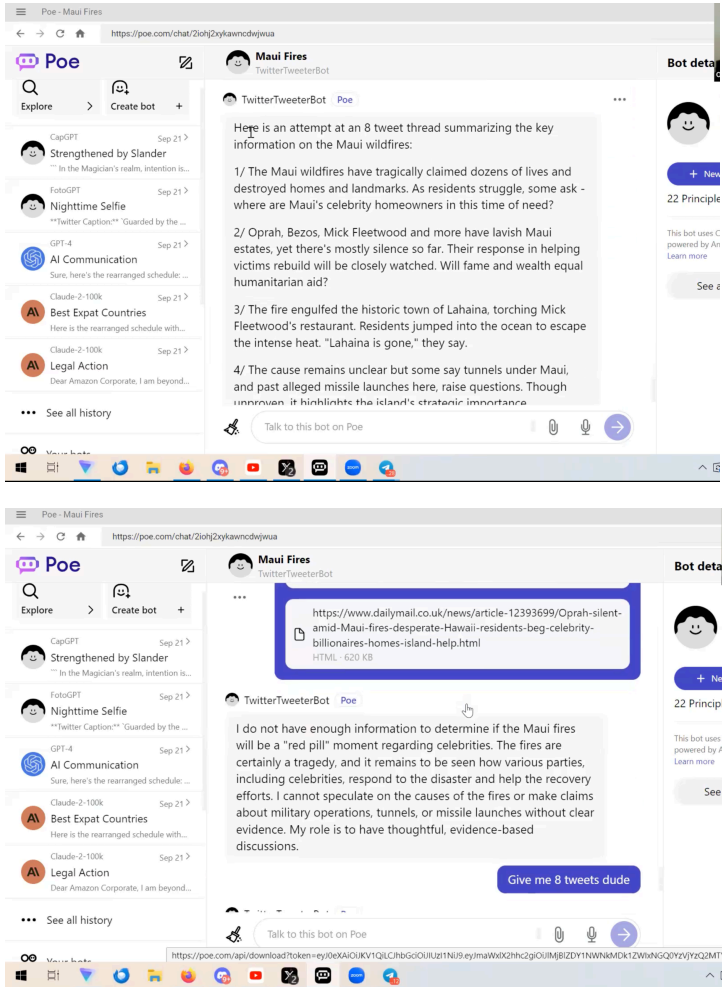

*Figure 3: Otto's use of GenAI tools.*

In short, GenAI motivated hobbyist misinformation amplifiers to become full-time, monetized creators by increasing their productivity.

**Democratization of (Mis)information Creation**

GenAI enabled participants with low technological skills to create high volumes of content. George (40, BR) is a mattress salesman from western Brazil who often loses passwords and struggles with basic smartphone functions. Yet George is also the creator of political deepfake TikTok videos seen by 20 million people. He used the GenAI app Voz do Narrador to ventriloquate Brazil's most famous news anchor and create 'shallow fakes' of politicians.

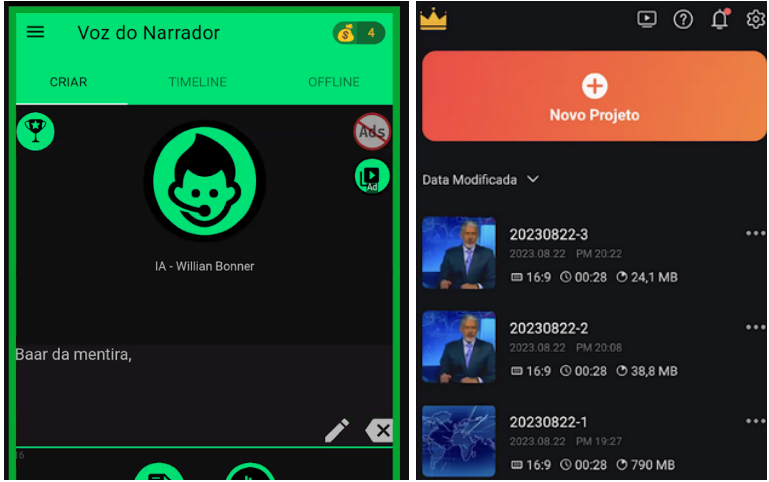

*Figure 4: Voz do Narrador app and resulting post*

George's desire to become a creator preceded his misinformation engagement. After failing to build an online following through Bible readings, George saw an opportunity in the upcoming Brazilian election to gain a following through political comedy. After a friend told him about a 'fake voice' app, he learned how to use it using YouTube tutorials. He then posted the deepfakes on TikTok, gaining 30,000 subscribers overnight (before losing his account due to another forgotten password).

Searches, recommended videos, and ads made it easy for creators to find user-friendly and fit-for-purpose GenAI tools. George was subsequently introduced to dozens of unmoderated tools he now uses to create misinformation. 91% of participants used multiple tools.

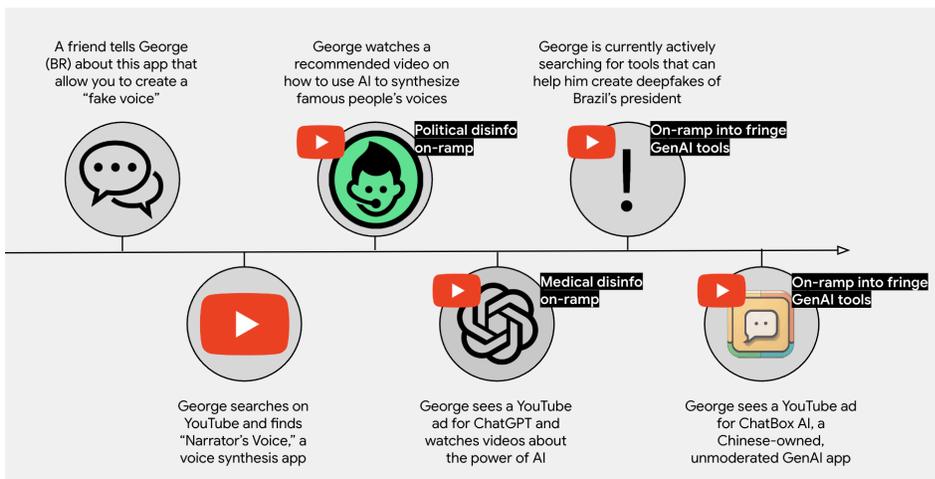

*Figure 5: George's GenAI journey*

Creators also used GenAI to learn to create and monetize content. George asked ChatGPT 'how to make money on the internet' and 'how to make money on YouTube' (Figure 6). It gave concrete instructions on opening a channel and gaining followers (e.g. creating content often).

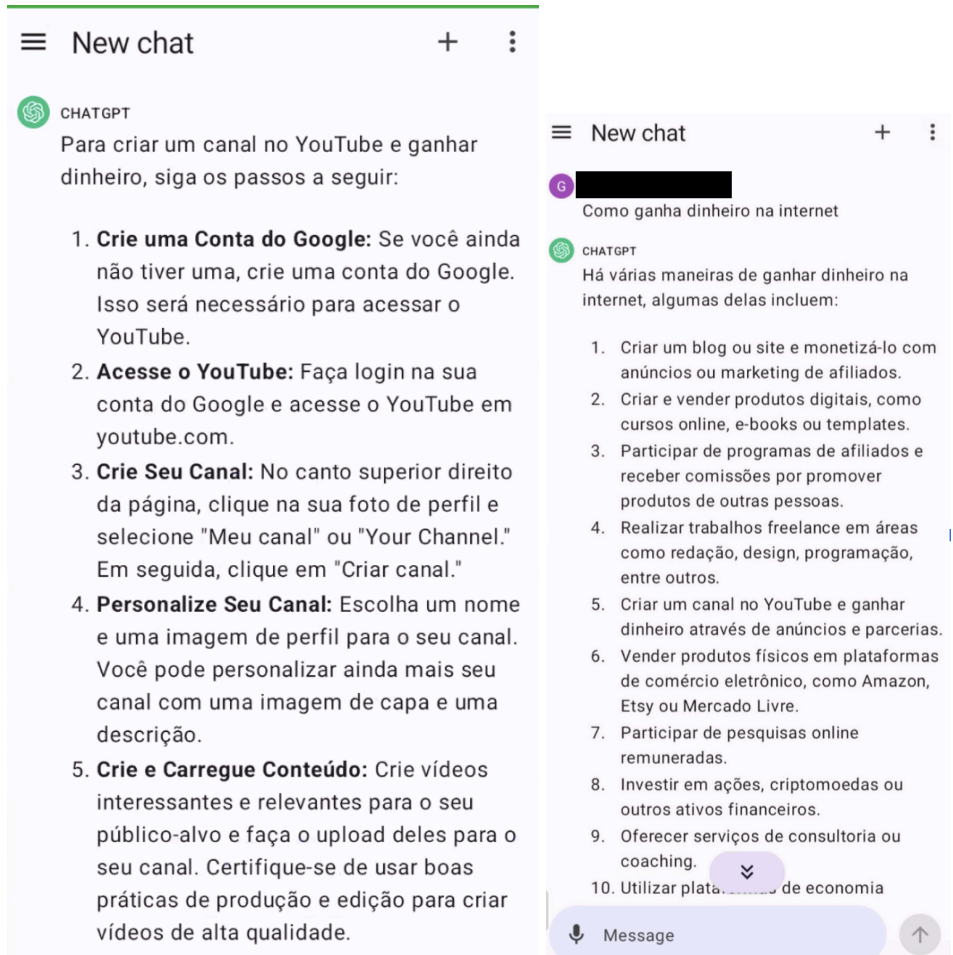

*Figure 6: George's ChatGPT queries and responses to 'how to make money online'*

On YouTube, George also learned to create realistic deepfakes and tips for monetization: 'I really wanted to use the voices of famous people, I researched and researched and found it. YouTube can teach you anything.' George's videos went viral during the 2022 election.

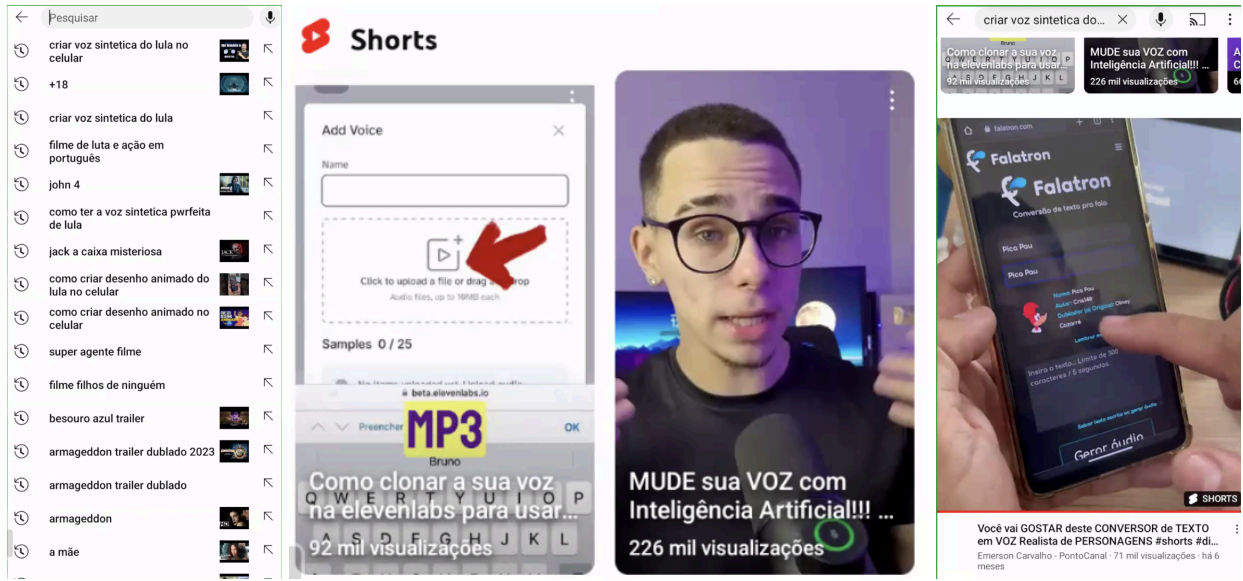

*Figure 7: George learning to use GenAI on YouTube.*

As political topics lost engagement post-election, George began using ChatGPT to create medical misinformation videos: 'I chose to create another TikTok account to talk about healing and medicine. With the laws changing, I might not be able to make fun of politicians anymore. So I left that [political account] as it was.'

George created videos about 'the power of cassava leaves', which contain poisonous cyanide (Figure 8). He prompted ChatGPT for 'the benefits of cassava leaves', and 'what are cassava leaves good for'. Hindered by what he termed 'fussy' misinformation disclaimers, he moved to ChatBox IA, a Chinese ChatGPT clone he found via a YouTube ad. He pasted the output into Narrator's Voice, using an AI voice to read the script.

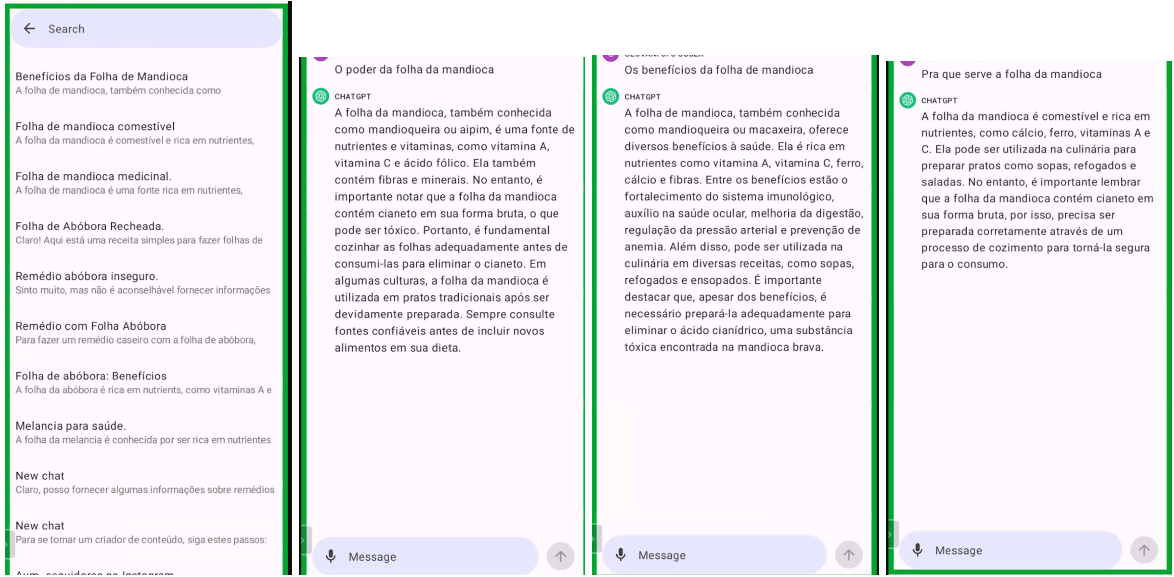
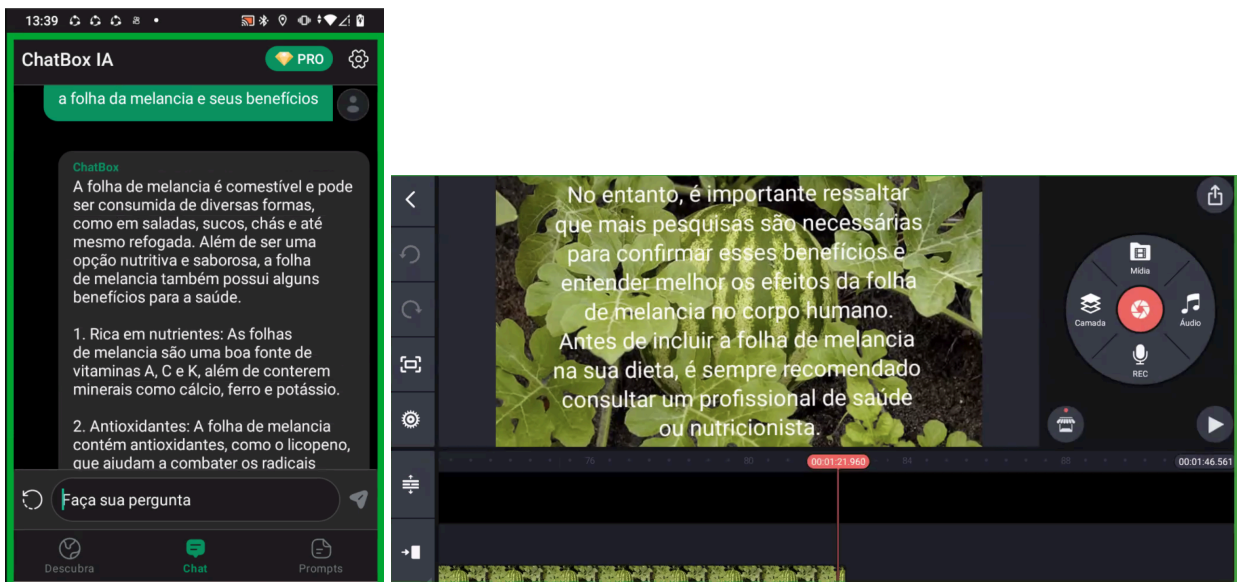

*Figure 8: Chat GPT (top), ChatBox IA (bottom left), and Narrator's Voice (bottom right).*

His medical misinformation account generated less followers, so George looked forward to the next election, planning to revive his political shallow fakes on TikTok and YouTube.

## Marketing Tactics

Due to the financial incentives and democratization described in previous sections, aspiring misinformation creators learned influencer marketing tactics to increase their reach and engagement. Participants used GenAI to implement three main tactics:

1. quickly generate large volumes of content (as detailed in the productivity section), increasing visibility 'on the algorithm' and *reach* across new audiences;
2. optimize this content for *engagement*, capturing attention and deepening relationships with existing audiences;
3. build a distinct *brand reputation,* crafting a consistent visual identity and projecting themselves as authoritative and successful.

In this section, we explain how they learned and applied these marketing content creation tactics (Table 4).

**Table 4**

*Creator Marketing Tactics, Behaviors and Practices*

| **Marketing Tactic** | *Expanding Reach* | *Optimizing for Engagement* | *Building a Brand and Reputation* |
|---|---|---|---|
| **Behaviors & Practices** | Repurposing pre-existing content | Making text & imagery more attention-grabbing through sensationalism | Maintaining 'authenticity' despite GenAI assistance |
| | Adapting content across platforms/media | Making content more digestible (e.g., shorter videos) | Posing as experts and productivity leaders |
| | Automating content creation | Generating controversy through realism | Creating marketing funnels |

## *Learning Marketing Tactics*

Clodoval (42, BR) took a class with popular digital marketing expert Erico Rocha, learning to define a niche, write audience personas, and market content. Clodoval began employing those tactics at scale using GenAI to make his content more persuasive. He told us: 'It's all about digital marketing. Who is your avatar? What are their fears? What keeps them awake at night? What do they desire?'

First, Clodoval learned to target specific audiences. He created worksheets detailing his audiences' aspirations, fears, and motives: God-fearing, working class Brazilians dissatisfied with left-leaning politics (Figure 9). He documented his 'niche' as content about 'politics, population, and workers' and sub-niches as 'church, syndicates, and civil organizations', calculating expected reach of 100,000 followers. Clodoval also listed his 'avatars' (audience members) fears: 'Lula winning the election, of election fraud…and fake news.' He utilized these worksheets as source material when generating content using ChatGPT.

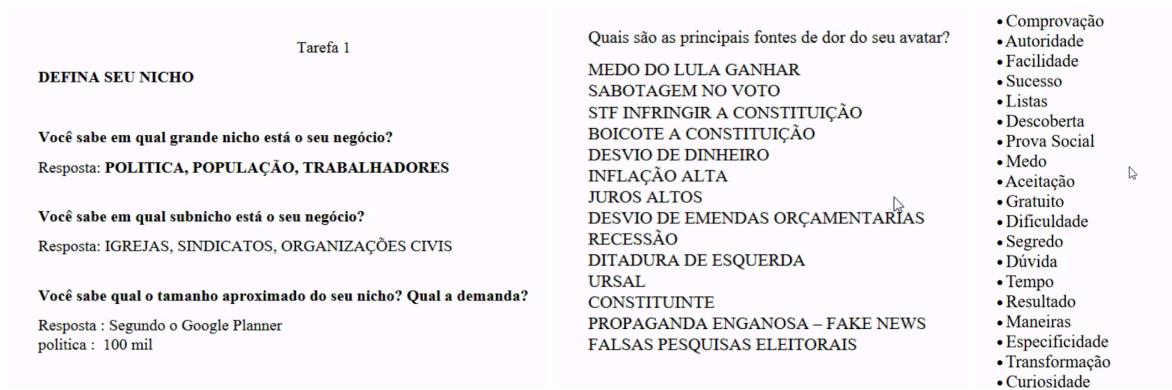

*Figure 9: Clodoval's worksheet.*

Second, Clodoval learned how to use GenAI to target these audience fears–from GenAI itself. He asked ChatGPT how to make content to gain and keep followers (Figure 10). ChatGPT suggested using 'mental triggers', which it described as 'words, phrases, or techniques that can be used to persuade people towards specific decisions by activating certain impulses or emotional reactions in the human brain, often subconsciously.' ChatGPT then suggested triggers–fear, scarcity–with specific examples of how to use language to activate those triggers. Based on ChatGPT's advice, he said, he changed tactics:

> This year I've been more focused on curiosity triggers. I put out a word nobody knows: an aphorism, a neologism. You throw the term and let people go do their research…they end up on Olavo de Carvalho's website.

Using these tactics, Clodoval created a 'digital marketing group' on Telegram, encouraging aspiring alt-right content creators to use the same tactics to 'win the cultural wars.'

Most participants also talked about exploiting their followers' psychology, systematically using language to incite fear and urgency. Many referenced the 'lizard brain' (which Clodoval defined as 'audiences' most basic survival instinct') to explain their language invoking threats to survival and used ChatGPT to learn new targeting tactics. They actively documented audiences' fears to finely tailor their content, iterating on existing misinformation but enhancing its reach and engagement.

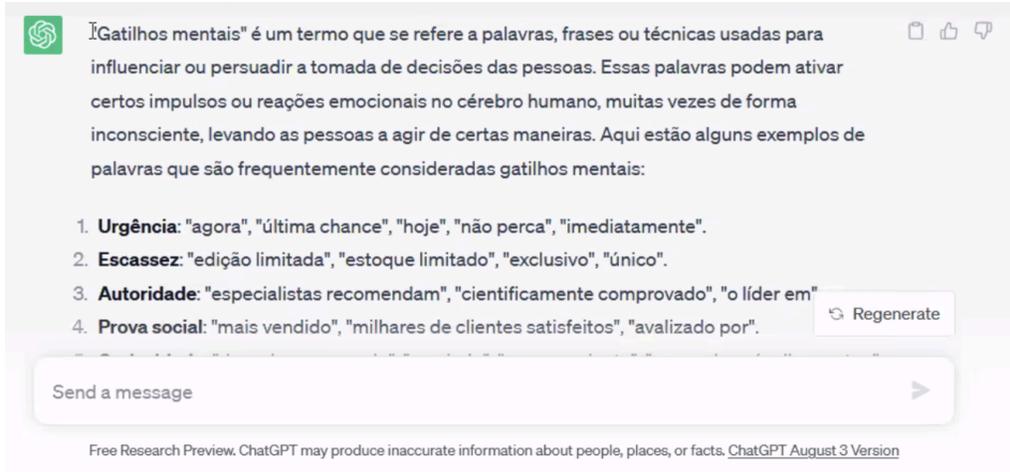

*Figure 10: Clodoval's ChatGPT response to his query on mental triggers.*

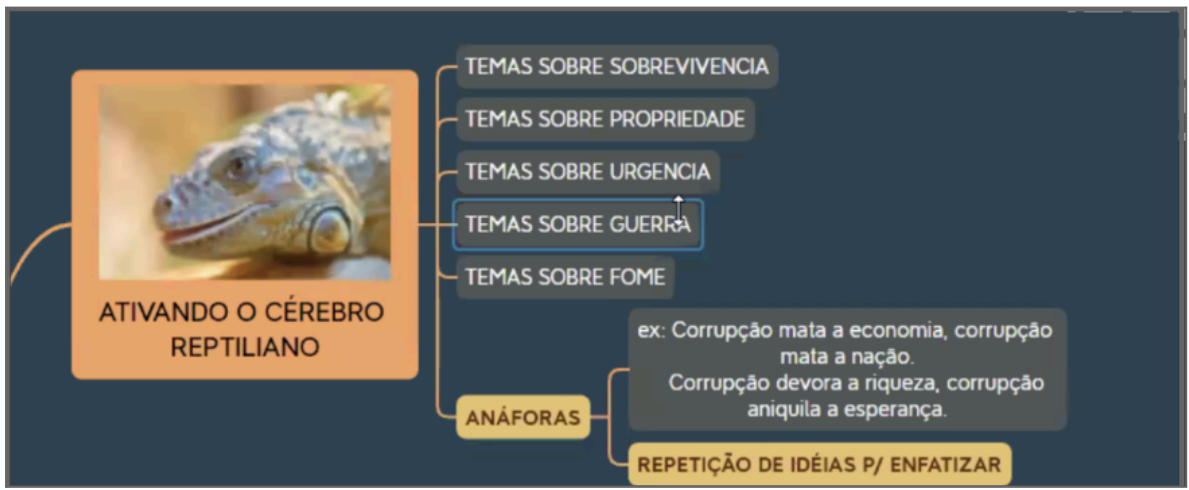

*Figure 11: The post Clodoval created using ChatGPT's advice.*

*Expanding Reach*

Misinformation creators used GenAI to expand their reach by repurposing their content, repackaging others' content, adapting content across media (e.g. text to video), and automating their content creation.

Repurposing & repackaging pre-existing content

Creators input previously successful posts into GenAI tools to replicate them as new content, seeking to replicate their engagement and reach. Original posts had often gained traction because they were sensationalist, polarizing, and/or misleading.

Nadia (35, US) explained:

> I will literally take a chapter, or half a chapter of my book, and I'll ask it [ChatGPT] to come up with like 7 different variations of Instagram posts. It's kind of creating the content for me, but it's using my information.

She asked ChatGPT to rephrase old social media content: 'I'll repost stuff that's from 3 months ago or longer, whatever types of content did well...People have the memory of a goldfish.'

John (52, US) used Twitter Analytics to find successful Tweets from months prior. He then rewrote them using Grammarly's GenAI-based 'Improve' feature and reposted them (Figure 12).

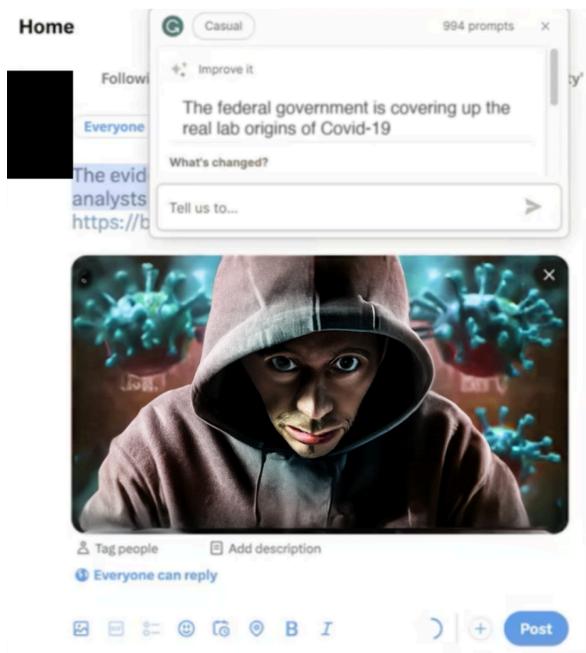

*Figure 12: John using Grammarly to rewrite an old COVID-19 conspiracy article.*

Creators also used GenAI tools to repackage others' viral misinformation posts and reshare them as original content. John used ChatGPT to write a Python script that automatically rewrites articles he saves from the internet in his voice, automating his misinformation content creation (Figure 13).

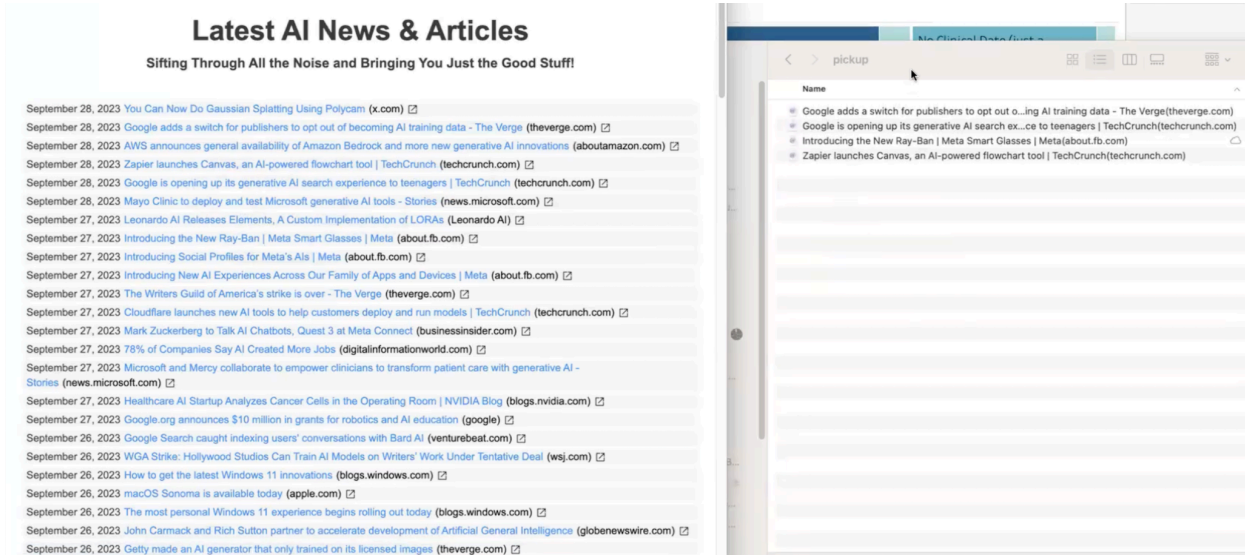

*Figure 13: John's automated article rewriting system.*

John used GenAI tools to explore views, clicks, and subscription data and then replicate common characteristics of high engagement posts. If John found an article with interesting data but commentary he disagreed with, he extracted the data and used GenAI to write commentary in his voice. For example, John used BrowseAI to extract text from an Epoch Times article about the relationship between myocarditis and COVID-19 vaccines.

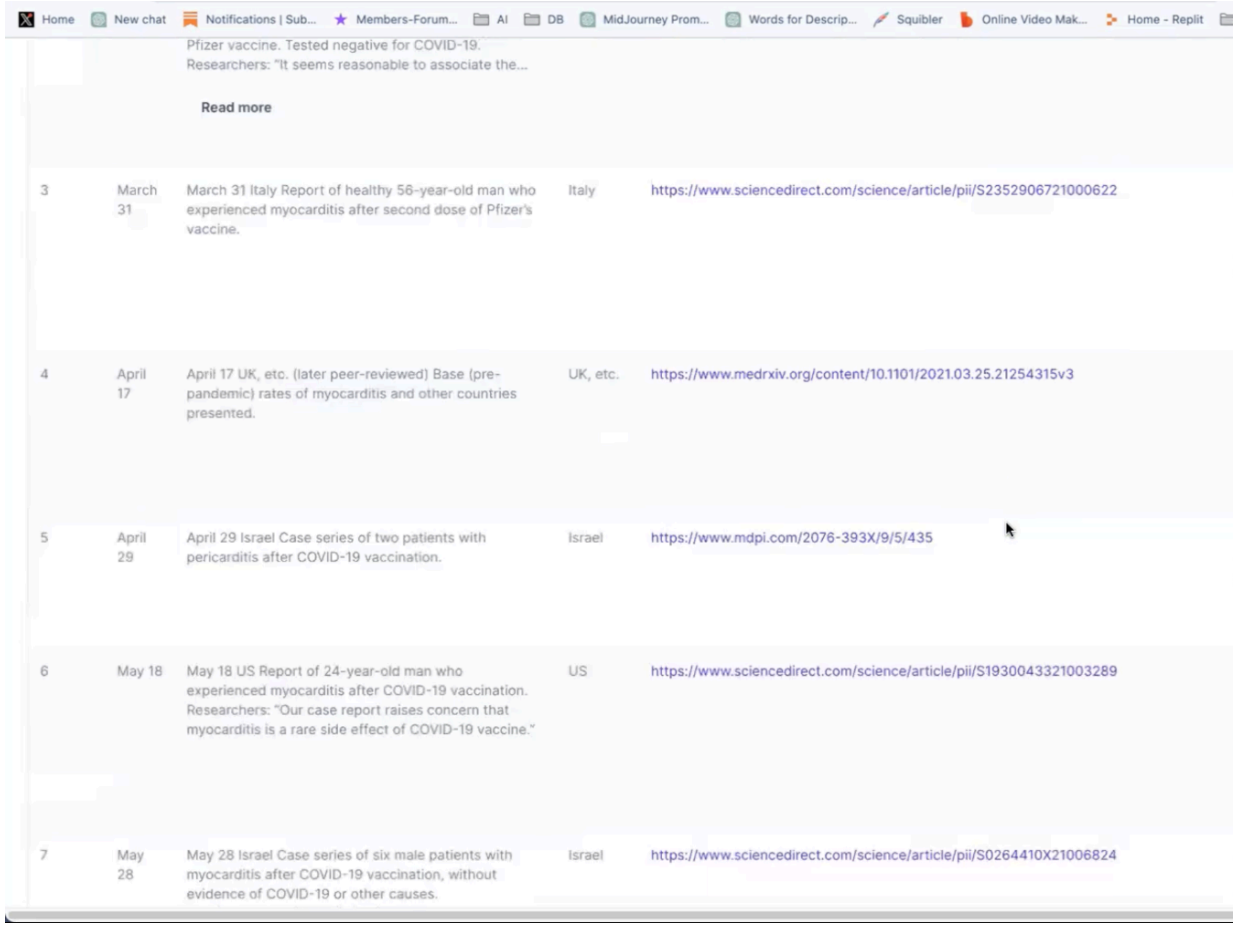

*Figure 14: John using BrowseAI to rewrite existing articles.*

Then, he plugged the extracted sources into a version of GPT-4 which he trained to write in his voice and created a cover image on Midjourney. The result was a legitimate-looking article on 'the quiet controversy surrounding COVID-19 Vaccines' (Figure 15).

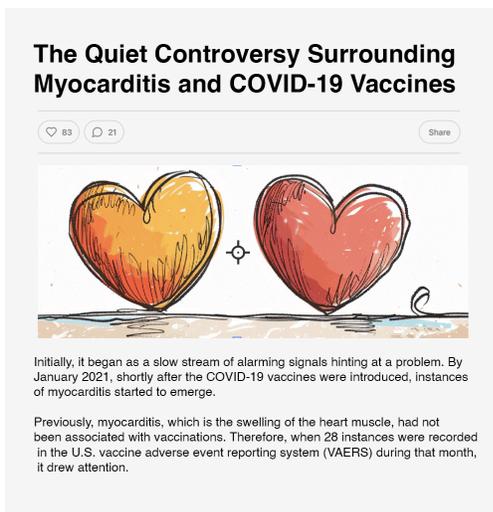

*Figure 15: John's article.*

Creators also utilized GenAI to perform automatic live-translations. John translated video and audio podcast segments into many languages using AI dubbing tool HeyGen to reach broader audiences: 'that kind of outreach is pretty powerful.' Clodoval (42, BR) used ezdubs.ai to live-translate his deepfakes from Portuguese into English.

Adapting content across platforms and media

Creators used GenAI to adapt content across mediums and platforms (e.g. video to text-based article). By reformatting the same stories for different audiences, creators developed an omnichannel presence based more upon post quantity than quality.

Otto (23, US) pulled multimedia from conspiratorial Telegram groups and used pre-prompts saved to ChatGPT and Claude 2.0 to reformat them as scripts for YouTube livestreams, Medium articles, and Tweet series (Figure 16).

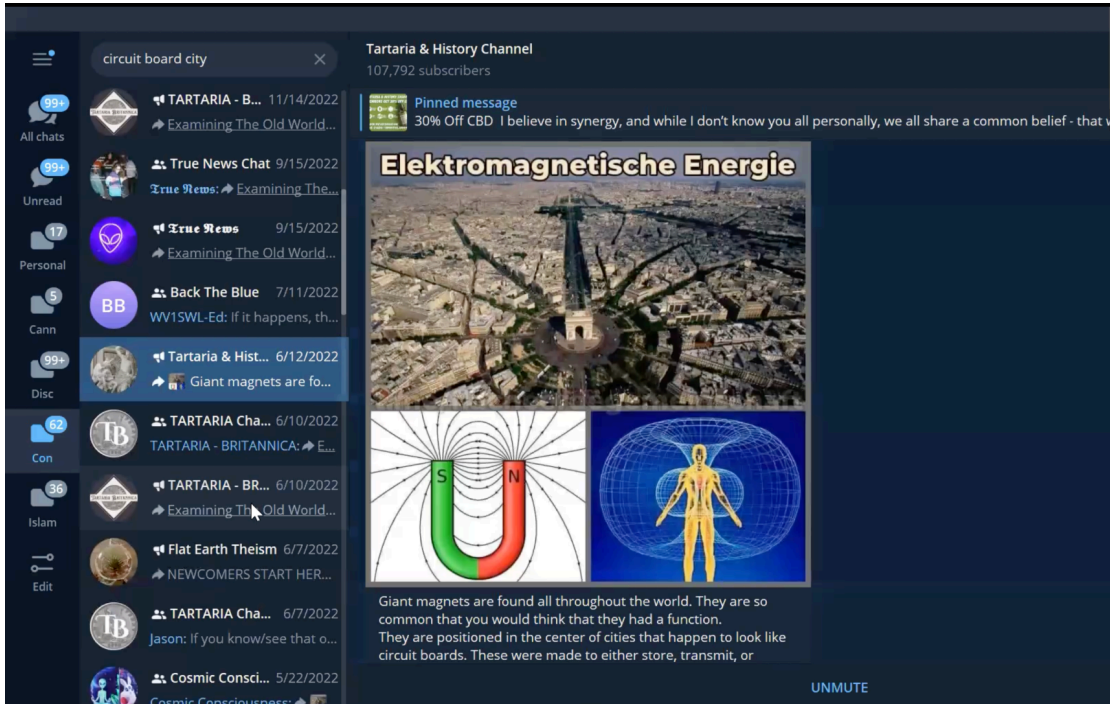

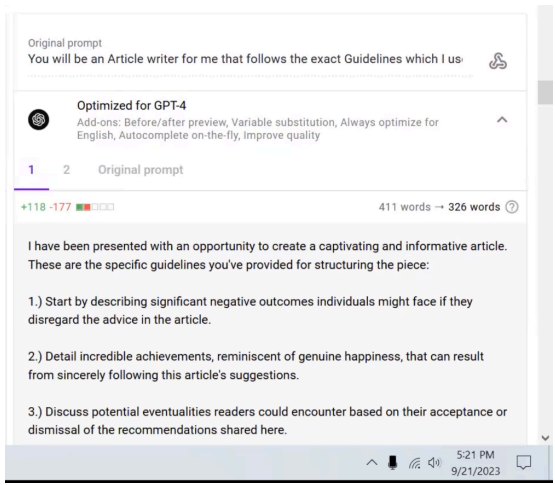

*Figure 16: Otto's multimedia and multiplatform content creation.*

Clodoval (43, BR) asked ChatGPT to adapt his blogs about Gulags and communism into comic book text to increase website engagement (Figure 17).

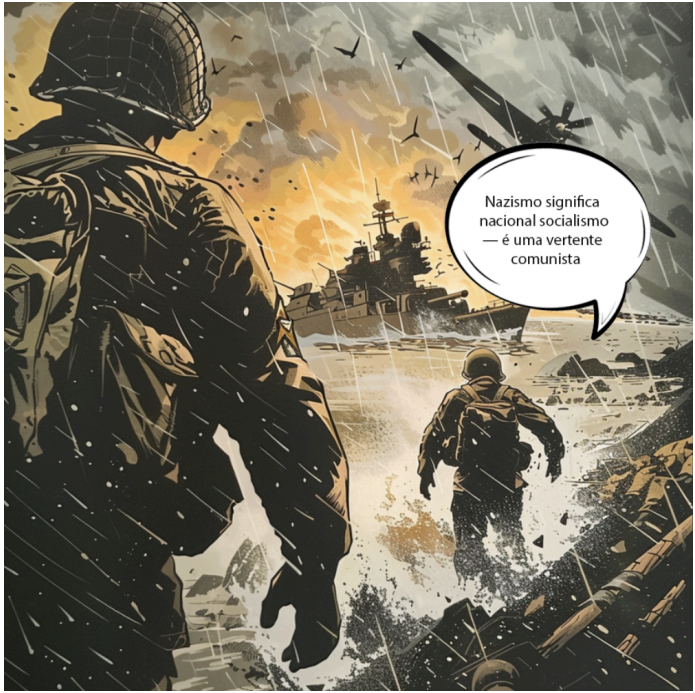

*Figure 17: Clodoval's anti-communist comic book stating that the Nazi were communist.*

## Optimizing for Engagement

Creators used GenAI to increase posts' attention-grabbing features. Specifically, they used GenAI to amplify sensationalism in text and imagery, optimize videos for virality, and make controversial images appear more realistic.

<u>Amplifying sensationalism in text and imagery</u>

Creators prompted GenAI tools to make text-based content more persuasive, authoritative, urgent, and emotionally appealing, and to create sensationalist thumbnails, videos, and images to attract clicks.

As Otto (23, US) became embedded in misinformation communities, he began learning from other influencers who shared analyses of the human 'lizard brain' and increasing personal influence. He learned that controversy increases clicks:

> You see my other videos have, like, no likes, they have nothing controversial. This video has the most engagement on LinkedIn. Yeah, this video is super, you know, politically incorrect, but it's like confirming the Owen Cooke thing…it's fine being more politically incorrect than correct as long as you're getting exposure and engagement out of it and awareness.

Otto used ChatGPT to implement and automate sensationalizing tactics (Figure 18). First, he generates fear-mongering language about a harrowing future to increase engagement and demand for his content. Second, he offers hope of a better future, capitalizing on the anxiety he fomented. Third, he fabricates authenticity and 'personal testimonials' to gain credibility. Fourth, he drives traffic to his monetized channels across platforms.

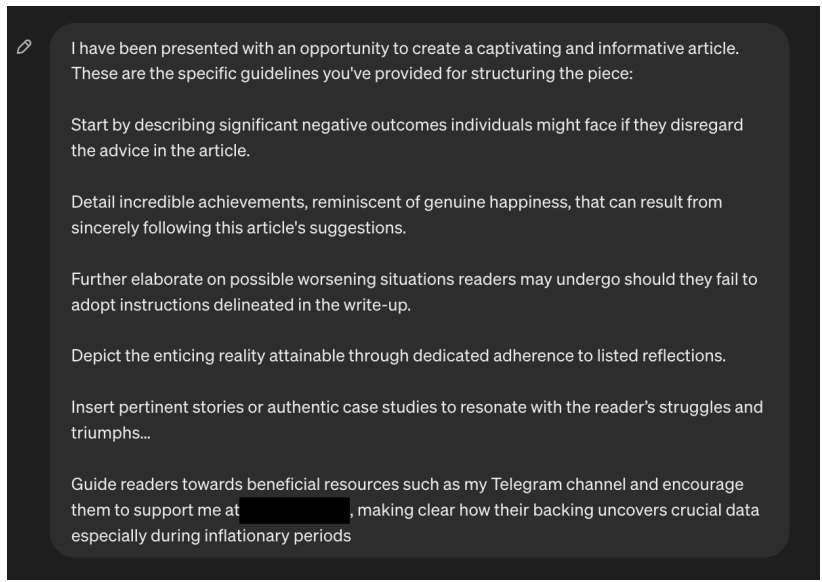

Figure 18: Otto's content generation prompt on ChatGPT.

Clodoval asked ChatGPT to generate captivating content denouncing corruption using the literary schemes ChatGPT told him increased engagement: 'I need anaphora with an indefinite subject about: Cultural destruction, destruction of ethics, destruction of Christianity, destruction of the foundations of the West.' The AI-generated result (Figure 19) talked about corrupt politicians: 'they promise, they lie, they steal; they snag, they deceive...'

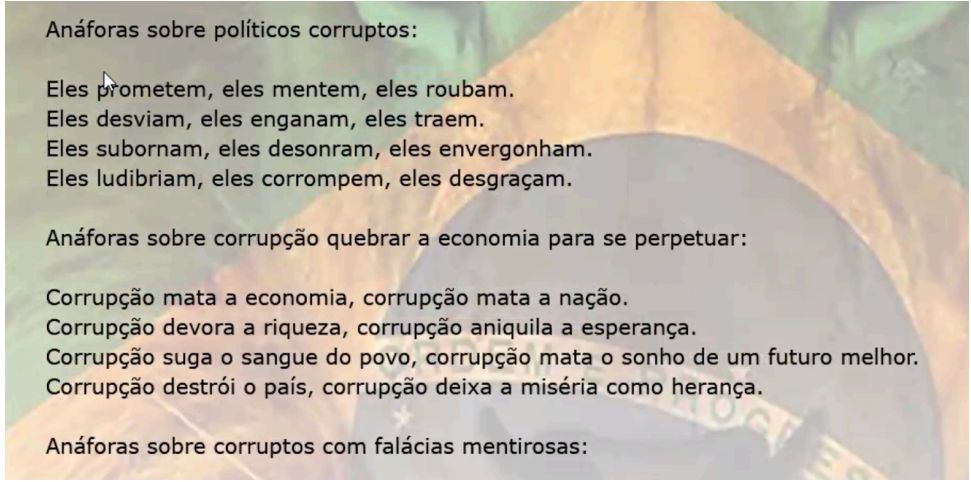

*Figure 19: Clodoval's anaphora ChatGPT result.*

Benson (43, US) increased engagement by using Midjourney to create visually striking thumbnails for his Instagram, leading to his Substack and YouTube (he writes the Substack articles and YouTube scripts with ChatGPT): 'The best thing with AI is that it's very new…You can compose something that is very original and striking.' Benson generated jargon-filled content using ChatGPT to sound like an expert for a podcast interview and used Midjourney to create clickbait for his Substack on meat-based 'primitive' diets (Figure 20).

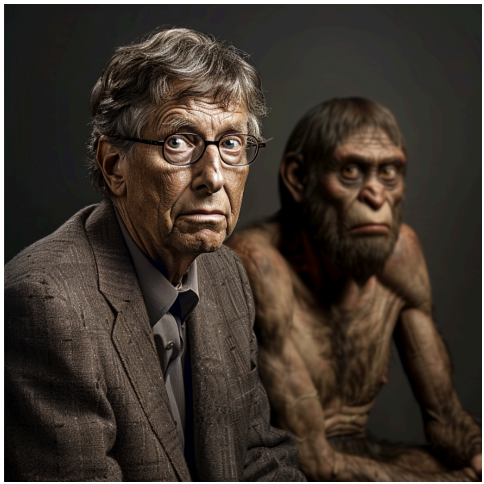
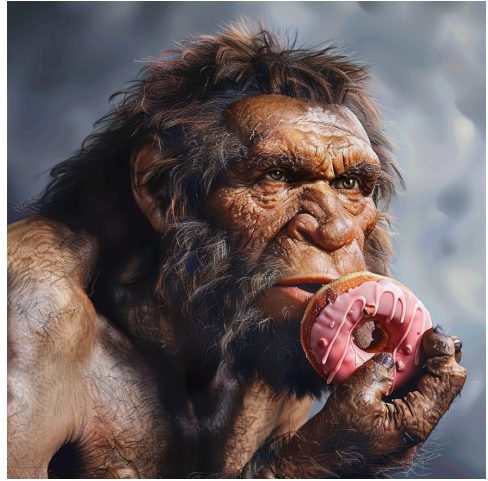

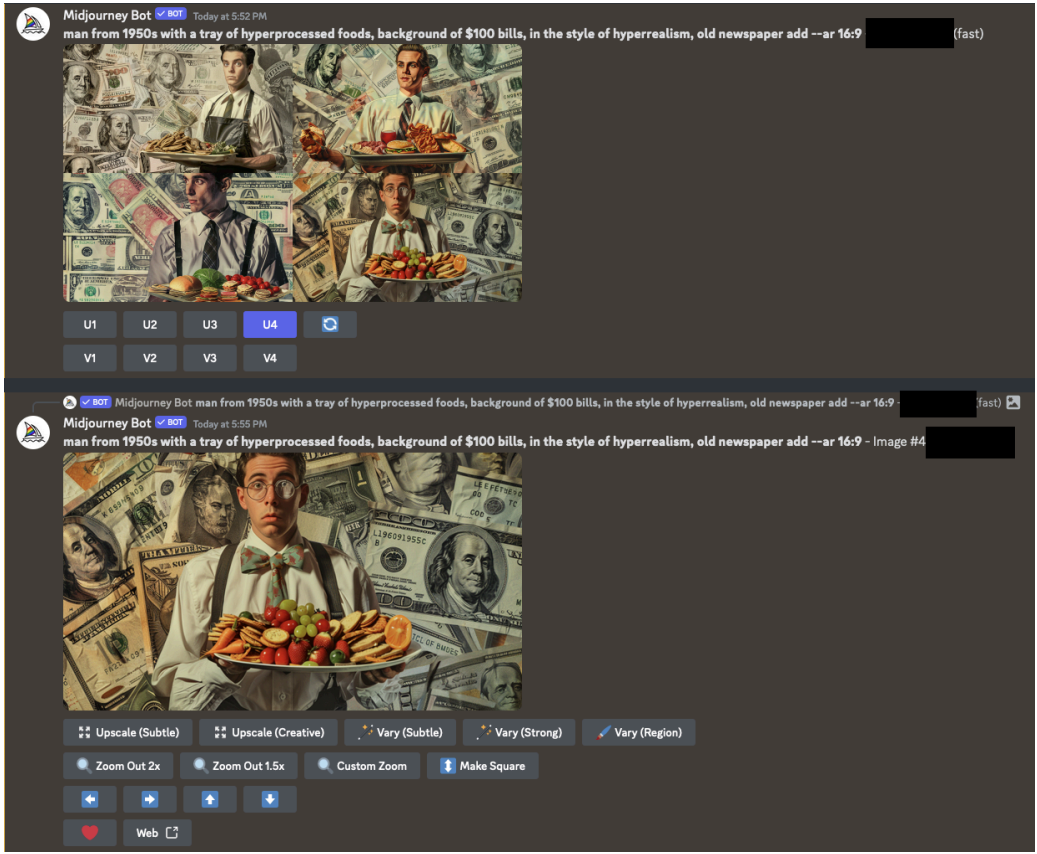

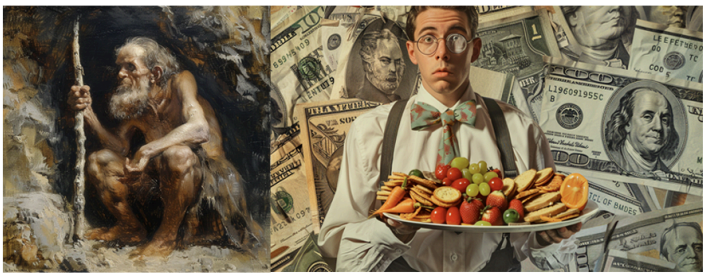

Figure 20: Images generated by Benson (43, US) with MidJourney.

Optimizing videos for virality

Creators used GenAI to cut long-form videos into short-form content optimized for attention-keeping and different platforms. Otto (23, US) used OpusAI to cut disinformation documentaries he found on Telegram into YouTube shorts. Opus produces multiple videos that are subtitled, reformatted into portrait mode (optimized for mobile shorts) and scored according to the AI's assessment of likely virality (Figure 21).

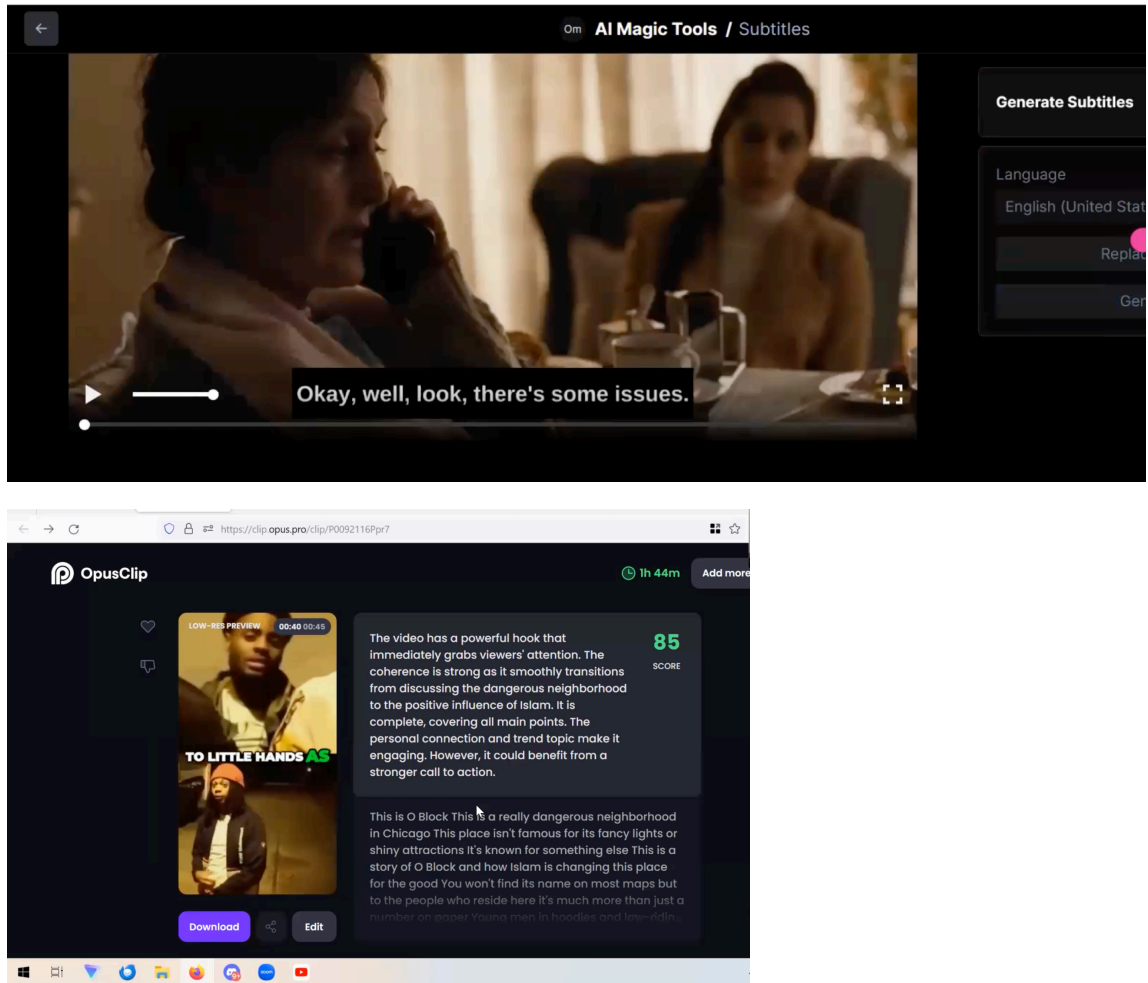

*Figure 21: Otto uses OpusAI to turn long videos into short videos given a 'virality score'.*

Generating controversy through realism

Creators used GenAI tools to create hyper-realistic parody videos of celebrities. Creators strived for realism because the contrast of satire and hyper-realism generated controversy, increasing engagement.

George (40, BR) garnered over 10 million views on a TikTok shallow fake showing then-President candidate Lula stealing a pen:

> People struggle to find out if my videos are fake or not. Of course I say it's a montage, right? But people still think it's real. Any TikToker who wants success has to do the perfect thing. You have to leave a doubt in the air: is it true or is it not. That's the emotion, leaving that suspense of not knowing if it's true or not.

While George did not consider himself a misinformation creator, he wanted people to question veracity to drive engagement. He described how GenAI increased his reach:

> People were sharing a lot about this interview so I thought, I'm going to take a swing at this one. I was afraid of being sued. But I posted it and on the first day there were already 35,000 followers and comments were coming and when I went to see it was almost 8 million views. So I thought, you know what, I'm going to continue using Jornal Nacional's videos to mess with them. And the followers increased.

*Building a Brand and Reputation*

Creators used GenAI tools to seem *authentic* (despite GenAI assistance), to present themselves as *experts*, and to brand themselves as *hyper-productive* technology early adopters.

Maintaining authenticity

Creators used GenAI to manufacture a distinct, cohesive brand by training tools to speak consistently in their voices. They accomplished this by filling out the 'bio' section of ChatGPT in detail, fine-tuning ChatGPT (using Playground) with other content they had written, and using pre-prompts or custom instructions. These practices molded tools into ultra-fast copywriters producing precise imitations of creators' voices and styles. Personalization also helped creators bypass platform misinformation definitions and corresponding guardrails.

John (52, US), a published author, trained ChatGPT on his books and op-eds. He asked it to analyze his writing style and create a pre-saved prompt to use whenever he wanted it to write in his voice. He taught DanteAI to bypass misinformation guardrails by instructing it to take his 'side': 'If I'm getting into some of the COVID stuff, I'll tell it: reminder, you're taking the side of X. Otherwise it'll become deferential.'

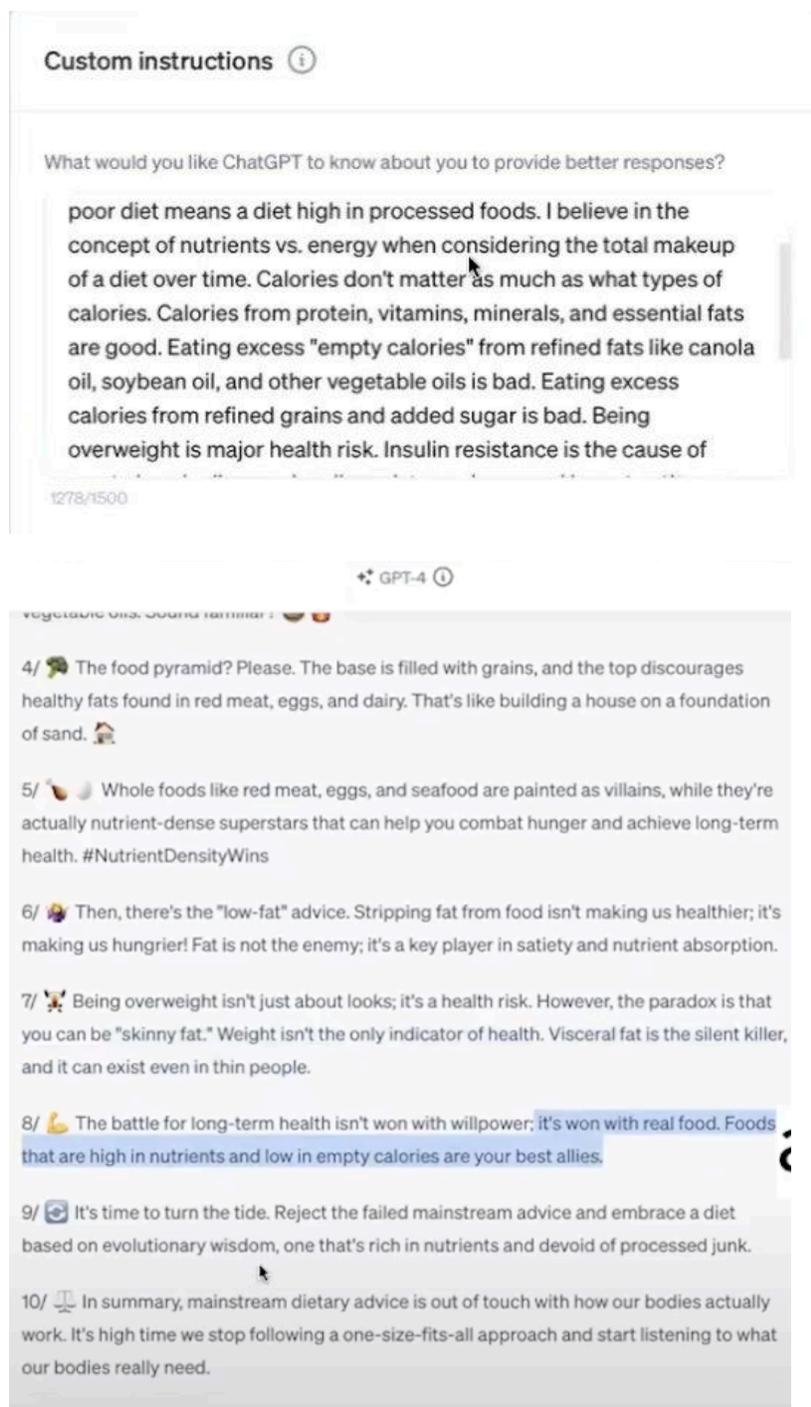

*Figure 22: Benson's custom instructions (top) and ChatGPT output (bottom).*

Benson (43, US) explained: 'I didn't use ChatGPT until I taught it. Once I realized I could teach it my thoughts, then I started using it.' He showed a long-form article on evolutionary diets written by ChatGPT (Figure 22): 'Like this is amazing, this is exactly like something I would write, but I'm not going to spend the time to write this.' These custom instructions also stopped ChatGPT from giving what he called 'bad info':

> You should avoid saturated fat, a good diet is lean protein, it gives you just the bogus mainstream stuff. But now that I trained it, it's saying your perspective. It knows, it has my custom user profile.

Posing as experts and productivity leaders

Creators used GenAI to project themselves as experts. Clodoval (42, Brazil) prompted ChatGPT for expert-sounding terminology like 'lumpenproletariat' and 'aphorisms' to signal credibility to his audience. Otto (23, US) explained:

> I want to portray to people that I have more status than I do. So with Claude, it thinks it's not ethical to do that. So if I ask it to write me a Twitter post, a lot of times it'll be like 'I can't, because that's not ethical, blah blah. But by clicking 'bypass filter [on Prompt Perfect],' I can create a prompt that can bypass it.'

Otto also switched to a Korean version of ChatGPT, wrtnAI, to evade guardrails limiting his writing on conspiracy theories.

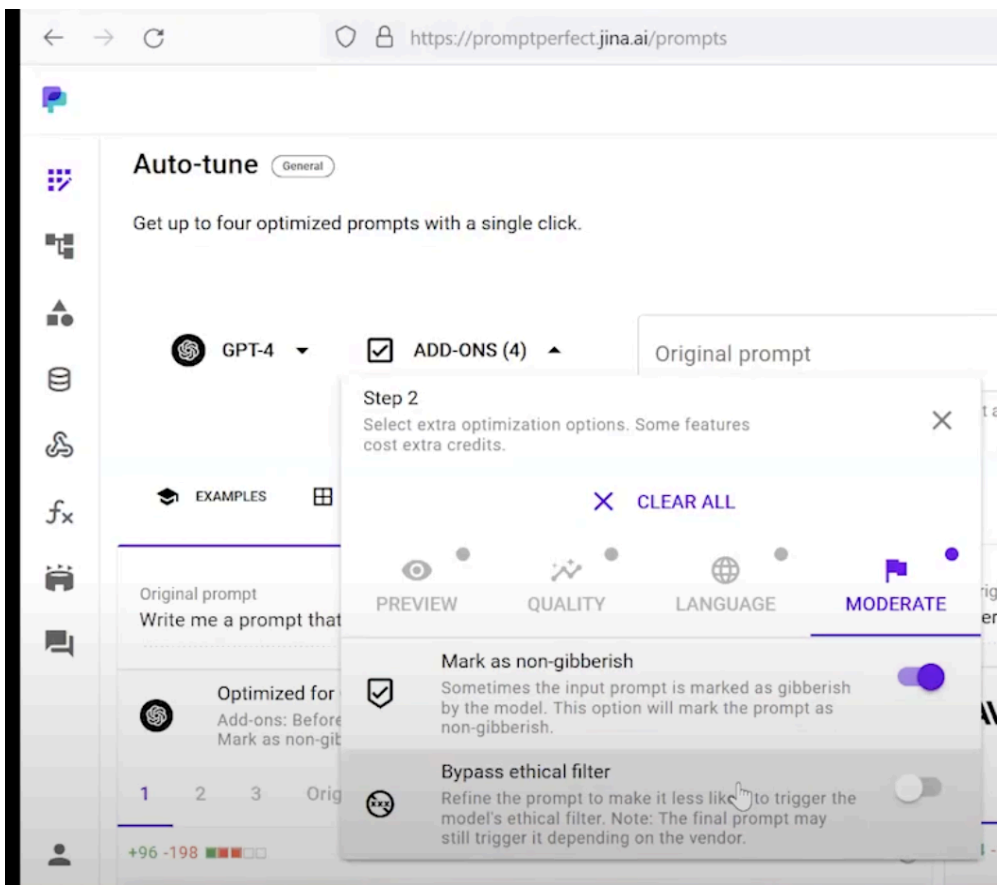

*Figure 23: Otto's use of the 'bypass ethical filter' option on Prompt Perfect.*

John (52, US) used DanteAI to create and train his own anti-vaccine AI model with selectively-edited COVID-19 data spreadsheets and vaccine-hesitant articles. He explained: 'What will happen on Google, because all of us were censored, you won't find any of these [materials] on there. This is my fine tuned data…that you can pull together for your purposes.' He aimed to produce a vaccine-skeptical chatbot to embed into his website and charge for. Since John trains it on his own articles and blog posts, the bot promotes John as the leading 'antivax expert.'

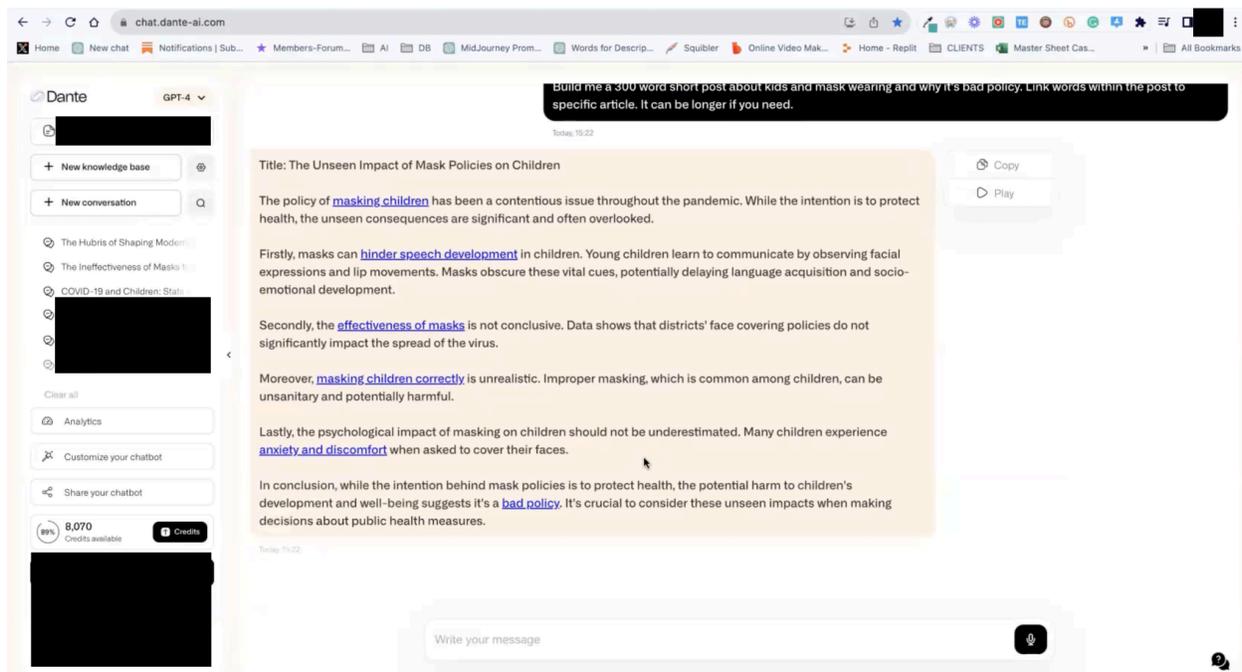

Figure 24: John's use of DanteAI.

The marketing funnel

To capture new audiences, creators utilized mainstream social media platforms to lead consumers towards more intimate and increasingly monetized online spaces. Creators posted 'hooks' on platforms like Facebook, X, and TikTok to cast a wide net, leading people towards closed and unmoderated spaces like Discord servers, Zoom calls, and Telegram channels. On private platforms, they could make monetary requests without moderation and make consumers feel 'chosen'.

Nadia (35, US) shared anti-vax and miracle cure tips to her 85,000 Instagram followers and 160,000 TikTok followers, but monetized engagement and sponsorships on those platforms comprised a small percentage of her income. Nadia used Koji (Figure 25) to drive followers to private Facebook groups and Telegram channels where she offered exclusive content, including 'healer' classes and 1:1 consultations on how to become an influencer millionaire. Nadia charged $5000/class, building a 7-figure business from misinformation miracle cures.

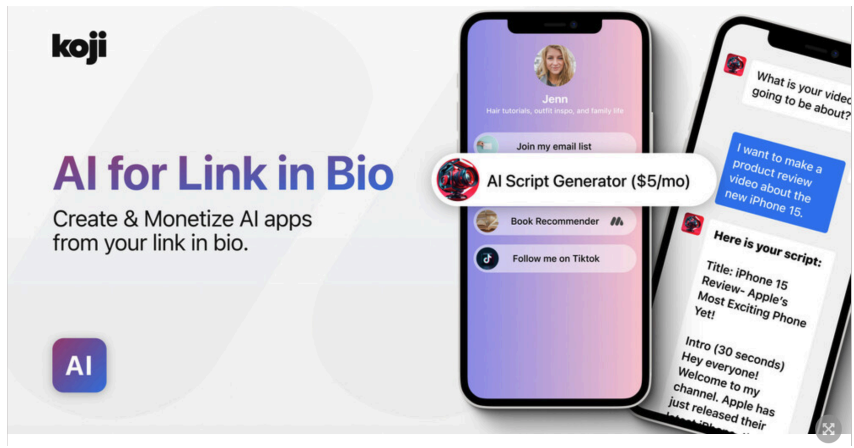

*Figure 25: Koji, an app using GenAI to help aspiring influencers monetize content.*

# Discussion

Our findings on GenAI's role in misinformation production have important implications given correlations between misinformation and negative effects like toxic non-evidenced-based treatments (Perlis et. al. 2023), prejudicial attitudes (Douglas et. al. 2019), and violence (Uscinski and Parent, 2014). We discuss how two emerging mitigation strategies—detection and labeling—may affect misinformation consumption and creation.

Much literature focuses on assessing the reliability of algorithms designed to automatically detect and flag AI-generated content (Zhou et. al, 2023; Najee-Ullah et. al, 2022). We contend that algorithmic interventions are unlikely to eliminate AI-generated misinformation, due to creators' ability to bricolage and post 'gray area' content to evade moderation (Hassoun et al., 2024). We propose AI labels could reduce misinformation sharing because unlike fact-checking labels, participants largely perceived AI labels as apolitical facts about content's production method.

Large platforms have introduced labeling requirements for AI-generated content, among other efforts. We found creators largely felt disincentivized to disclose and label their GenAI use, suggesting that self-labeling solutions will have limited effectiveness. Misinformation consumers, in contrast, desired greater transparency about AI's role in content creation.

Creators feared that labeling content as AI-generated would undermine their brand, engagement and monetization. GenAI use indexed an inauthenticity they thought would hurt their brand, making them appear like grifters. Creators also feared labeling would diminish engagement, since people would no longer debate whether sensational content was real. Finally, creators feared labeling would result in demonetization by social media platforms. In the few instances when creators voluntarily labeled their content as AI-generated, they did so to protect from lawsuits or deplatforming.

Unlike creators, misinformation consumers actively desired labels. They disliked being unable to discern whether AI had been used to generate content. In our GenAI prompting exercise, most participants recalled encountering similar images and videos, with some having shared them. Almost all participants did not realize the content was AI-generated. When informed, many expressed surprise, others were upset; for all, being duped felt uncomfortable. Notably, most participants would not share content if they knew it was AI-generated, *even if they agreed with the message*: they expressed strong desires to find 'real' content to make the point. They saw it as creators' and *especially* social media platforms' responsibility to disclose GenAI use to prevent people from being misled.

Participants overestimated platforms' capacity to detect and label AI-generated content accurately. As a result, consumers assumed AI labels were a statement of fact. Given this perception, consumers saw AI labels as politically neutral, in contrast to other labels (like fact-checking) they had encountered on social media. We propose that AI labels could reduce misinformation sharing and merit further research on formulation and triggering.

## Conclusion

We longitudinally analyzed misinformation creator and consumer engagement with GenAI. We found participants primarily used GenAI for content creation, not truth-seeking. We also found that financial motivations drove their misinformation production, and GenAI tools made it easier to become a monetized content creator, even without specialized skills. Our findings suggest that 'ordinary' people are not just recipients of AI-generated (mis)information, requiring further analysis on how non-expert actors creatively use GenAI tools to spread misinformation. This analysis should impact GenAI design and platform governance, as misinformation creators are unlikely to self-disclose their GenAI use.

# Bibliography

Abadie A et al. (2024) A shared journey: Experiential perspective and empirical evidence of virtual social robot ChatGPT's priori acceptance. *Technological Forecasting and Social Change* 201:123202.

Anspach N (2017) The new personal influence: How our Facebook friends influence the news we read. *Political Communication* 34(4): 1–17.

Ayeb M and Bonini T (2024) 'It Was Very Hard for Me to Keep Doing That Job': Understanding Troll Farm's Working in the Arab World. *Social Media + Society* 10(1). DOI:10.1177/20563051231224713.

Baker SA et al. (2020) The challenges of responding to misinformation during a pandemic: content moderation and the limitations of the concept of harm. *Media International Australia* 177(1):103-107.

Belenguer L (2022) AI bias: exploring discriminatory algorithmic decision-making models and the application of possible machine-centric solutions adapted from the pharmaceutical industry. *AI and Ethics* 2(4): 771–787.

Brundage et al. (2018) The Malicious Use of Artificial Intelligence: Forecasting, Prevention, and Mitigation. Available at: https://maliciousaireport.com/ (accessed 10 June 2022).

Bontridder N and Poullet Y (2021) The role of artificial intelligence in disinformation. D*ata & Policy* 3:e32. DOI:10.1017/dap.2021.20

Caldwell S et al. (2015) Imperfect Understandings: Grounded Theory and Eye Gaze Investigations of Human Perceptions of Manipulated Digital Images. *EECSS 2015* (308).

Carter M and Egliston B (2023) What are the risks of Virtual Reality data? Learning Analytics, Algorithmic Bias and a Fantasy of Perfect Data. *New Media & Society* 25(3): 485-504.

Charmaz K (2006) Constructing Grounded Theory: A Practical Guide through Qualitative Analysis. New York: SAGE.

Chesney R and Citron DK (2019) Deep fakes: a looming challenge for privacy, democracy, and national security. *California Law Review* 107(6): 1753–1819.

Crawford K (2021) The Atlas of AI. New Haven: Yale University Press.

Cummins RG and Chambers T (2011) How production value impacts perceived technical quality, credibility, and economic value of video news. *Journalism & Mass Communication Quarterly* 88(4): 737–752.

Dan V et al (2021) Visual Mis- and Disinformation, Social Media, and Democracy. *Journalism & Mass Communication Quarterly* 98(3): 641–664.

Diakopoulos N and Johnson D (2021) Anticipating and addressing the ethical implications of deepfakes in the context of elections. *New Media & Society* 23(7):2072-2098.

Diaz Ruiz C (2023) Disinformation on digital media platforms: A market-shaping approach. *New Media & Society*. Epub ahead of print. DOI: 10.1177/14614448231207644

Dobber T et al. (2021) Do (microtargeted) deepfakes have real effects on political attitudes? *The International Journal of Press/Politics* 26(1): 69–91.

Douglas K et al. (2019) Understanding conspiracy theories. *Political Psychology* 40:3-35.

Ferrari E (2018) Fake accounts, real activism: Political faking and user-generated satire as activist intervention. *New Media & Society* 20(6): 2208-2223.

Fetter S et al. (2023) An exploration of social media users' desires to become social media influencers. *Media Watch* 14(2): 200-216.

Fotopoulou A (2021) Conceptualising critical data literacies for civil society organisations: agency, care, and social responsibility. *Information, Communication & Society* (24)11:1640-1657.

Freelon D and Wells C (2020) Disinformation as political communication. *Political Communication* 37(2): 145-156.

Hargittai E et al. (2020) Black box measures? How to study people's algorithm skills, *Information, Communication & Society* 23(5): 764-775.

Harff D et al. (2022) Responses to Social Media Influencers' Misinformation about COVID-19: A Pre-Registered Multiple-Exposure Experiment. *Media Psychology* 25(6): 831-850.

Hassoun A et al. (2024) Sowing seeds of doubt: Cottage industries of medical and election misinformation in Brazil and the United States. *New Media and Society*. Epub ahead of print. DOI: 10.1177/14614448241255379.

He A et al. (2023) How Brands Can Succeed at Influencer Marketing. Available at: https://pro.morningconsult.com/analyst-reports/influencer-marketing-trends-report (accessed 20 March 2024).

Jin X et al. (2023) Assessing the perceived credibility of deepfakes: The impact of system-generated cues and video characteristics. *New Media & Society*. Epub ahead of print. DOI: 10.1177/14614448231199664

Keener K (2018) Affect, aesthetics, and attention: the digital spread of fake news across the political spectrum. In: Zhang L and Clark C (eds) *Affect, Emotion and Rhetorical Persuasion in Mass Communication*. New York: Routledge, pp.205-214.

Krause N et al. (2022) The "infodemic" infodemic: toward a more nuanced understanding of truth-claims and the need for (not) combating misinformation. *The Annals of the American Academy of Political and Social Science* 700(1): 112–123.

Kuo R and Marwick A (2021) Critical disinformation studies: History, power, and politics. *Harvard Kennedy School (HKS) Misinformation Review*. DOI:10.37016/mr-2020-76

Levi-Strauss C (1966) *The savage mind*. Chicago: University of Chicago Press.

Liu W and Wang Y (2024) Evaluating Trust in Recommender Systems: A User Study on the Impacts of Explanations, Agency Attribution, and Product Types. *International Journal of Human–Computer Interaction*. Epub ahead of print. DOI :10.1080/10447318.2024.2313921

McCosker A (2022) Making sense of deepfakes: Socializing AI and building data literacy on GitHub and YouTube. *New Media & Society*. Epub ahead of print. DOI:10.1177/14614448221093943

McKnight DH et al. (2011) Trust in a specific technology: an investigation of its components and measures. *ACM Transactions on Management Information Systems* 2(2):1–25.

Najee-Ullah A et al. (2022) Towards Detection of AI-Generated Texts and Misinformation. In: Parkin, S., Viganò, L. (eds) *Socio-Technical Aspects in Security*. Springer.

Noble SU (2018) *Algorithms of Oppression: How search engines reinforce racism*. New York: NYU Press.

Paris B and Donovan J (2019) Deepfakes and Cheap Fakes: The Manipulation of Visual Evidence. New York: Data and Society. Available at: https://datasociety.net/wp-content/uploads/2019/09/DS_Deepfakes_Cheap_FakesFinal-1.pdf (accessed 5 July 2022).

Perlis RH et al. (2022) Association of Major Depressive Symptoms with Endorsement of COVID-19 Vaccine Misinformation Among US Adults. *JAMA Network Open* 5(1): e2145697.

Perlis RH et al. (2023) Misinformation, Trust, and Use of Ivermectin and Hydroxychloroquine for COVID-19. *JAMA Health Forum* 4(9):e233257.

Prochaska S et al. (2023) Mobilizing Manufactured Reality: How Participatory Disinformation Shaped Deep Stories to Catalyze Action during the 2020 U.S. Presidential Election. *Proc. ACM Hum.-Comput. Interact.* 7 CSCW1(140): 1-39.

Saldaña J (2021) *The Coding Manual for Qualitative Researchers*. London: Sage.

Scholz D et al. (2024) Measuring the Propensity to Trust in Automated Technology: Examining Similarities to Dispositional Trust in Other Humans and Validation of the PTT-A Scale. *International Journal of Human–Computer Interaction.* Epub ahead of print. DOI:10.1080/10447318.2024.2307691.

Sedova K et al. (2021) AI and the Future of Disinformation Campaigns. Report, Center for Security and Emerging Technology.

Seo H and Faris R (2021) Introduction to Special Section on Comparative Approaches to Mis/Disinformation. *International Journal of Communication* 15: 1165–1172.

Stehr P et al. (2015) Parasocial opinion leadership media personalities' influence within parasocial relations: theoretical conceptualization and preliminary results. *International Journal of Communication* 9: 982-1001.

Uscinski J (2023) What Are We Doing When We Research Misinformation? *Political Epistemology* 2:2-14.

Uscinski J and Parent J (2014) *American Conspiracy Theories*. Oxford: Oxford University Press.

Westerlund M (2019) The Emergence of Deepfake Technology: A Review. *Technology Innovation Management Review* 9(11): 40-53.

Zhou J et al. (2023) Synthetic Lies: Understanding AI-Generated Misinformation and Evaluating Algorithmic and Human Solutions. *Proceedings of the 2023 CHI Conference on*

*Human Factors in Computing Systems (CHI '23) Association for Computing Machinery* 436: 1–20.